\documentstyle[10pt,epsf,epsfig,wrapfig,dp_delphititle]{dp_delphi}
\makeindex
\def\DpPaperGroup{EP}
\def\DpPaperRef{2000-150 }
\def\DpDate{24 February 2000} 
\def\DpAuthors{DELPHI Collaboration}
\def\DpSubmit{(Eur. Phys. J. C19(2001)15)}
\def\DpTitle{{Study of Dimuon Production in Photon-Photon 
Collisions and Measurement of QED Photon Structure Functions at LEP}}

\begin{document}
\makeatletter
\newcount\@tempcntc
\def\@citex[#1]#2{\if@filesw\immediate\write\@auxout{\string\citation{#2}}\fi
  \@tempcnta\z@\@tempcntb\m@ne\def\@citea{}\@cite{\@for\@citeb:=#2\do
    {\@ifundefined
       {b@\@citeb}{\@citeo\@tempcntb\m@ne\@citea\def\@citea{,}{\bf ?}\@warning
       {Citation `\@citeb' on page \thepage \space undefined}}%
    {\setbox\z@\hbox{\global\@tempcntc0\csname b@\@citeb\endcsname\relax}%
     \ifnum\@tempcntc=\z@ \@citeo\@tempcntb\m@ne
       \@citea\def\@citea{,}\hbox{\csname b@\@citeb\endcsname}%
     \else
      \advance\@tempcntb\@ne
      \ifnum\@tempcntb=\@tempcntc
      \else\advance\@tempcntb\m@ne\@citeo
      \@tempcnta\@tempcntc\@tempcntb\@tempcntc\fi\fi}}\@citeo}{#1}}
\def\@citeo{\ifnum\@tempcnta>\@tempcntb\else\@citea\def\@citea{,}%
  \ifnum\@tempcnta=\@tempcntb\the\@tempcnta\else
   {\advance\@tempcnta\@ne\ifnum\@tempcnta=\@tempcntb \else \def\@citea{--}\fi
    \advance\@tempcnta\m@ne\the\@tempcnta\@citea\the\@tempcntb}\fi\fi}
 
\makeatother
\begin{titlepage}
\pagenumbering{roman}
\CERNpreprint{\DpPaperGroup}{\DpPaperRef} 
\date{{\small\DpDate}} 
\title{\DpTitle} 
\address{\DpAuthors} 
\begin{shortabs} 
\noindent
%
\noindent

Muon pair production in the process $e^+e^-\to e^+e^-\mu^+\mu^-$
is studied using the data taken at LEP1 ($\sqrt{s}\simeq m_Z$) 
with the DELPHI detector during the years 1992-1995. The corresponding 
integrated 
luminosity is 138.5~pb$^{-1}$. The QED predictions have been tested over
the whole $Q^2$ range accessible at LEP1 (from several GeV$^2/c^4$ to 
several hundred GeV$^2/c^4$) by comparing experimental distributions 
with distributions resulting from Monte Carlo simulations using various 
generators.
Selected events are used to extract the leptonic photon structure
function $F_2^\gamma$.
Azimuthal correlations are used to obtain information on additional structure 
functions, $F_A^\gamma$ and $F_B^\gamma$, which originate from 
interference terms of the scattering amplitudes. The measured ratios
$F_A^\gamma/F_2^\gamma$ and $F_B^\gamma/F_2^\gamma$ are significantly 
different from zero and consistent with QED predictions.
\end{shortabs}
\vfill
\begin{center}
\DpSubmit \ \\ 
\end{center}
\vfill
\clearpage
\headsep 10.0pt
\addtolength{\textheight}{10mm}
\addtolength{\footskip}{-5mm}
\begingroup
%
\newcommand{\DpName}[2]{\hbox{#1$^{\ref{#2}}$},\hfill}
\newcommand{\DpNameTwo}[3]{\hbox{#1$^{\ref{#2},\ref{#3}}$},\hfill}
\newcommand{\DpNameThree}[4]{\hbox{#1$^{\ref{#2},\ref{#3},\ref{#4}}$},\hfill}
\newskip\Bigfill \Bigfill = 0pt plus 1000fill
\newcommand{\DpNameLast}[2]{\hbox{#1$^{\ref{#2}}$}\hspace{\Bigfill}}
%
\footnotesize
\noindent
\DpName{P.Abreu}{LIP}
\DpName{W.Adam}{VIENNA}
\DpName{T.Adye}{RAL}
\DpName{P.Adzic}{DEMOKRITOS}
\DpName{I.Ajinenko}{SERPUKHOV}
\DpName{Z.Albrecht}{KARLSRUHE}
\DpName{T.Alderweireld}{AIM}
\DpName{G.D.Alekseev}{JINR}
\DpName{R.Alemany}{VALENCIA}
\DpName{T.Allmendinger}{KARLSRUHE}
% ...
\normalsize
\endgroup
\titlefoot{Department of Physics and Astronomy, Iowa State
     University, Ames IA 50011-3160, USA
    \label{AMES}}
\titlefoot{Physics Department, Univ. Instelling Antwerpen,
     Universiteitsplein 1, B-2610 Antwerpen, Belgium \\
     \indent~~and IIHE, ULB-VUB,
     Pleinlaan 2, B-1050 Brussels, Belgium \\
     \indent~~and Facult\'e des Sciences,
     Univ. de l'Etat Mons, Av. Maistriau 19, B-7000 Mons, Belgium
    \label{AIM}}
\titlefoot{Physics Laboratory, University of Athens, Solonos Str.
     104, GR-10680 Athens, Greece
    \label{ATHENS}}
\titlefoot{Department of Physics, University of Bergen,
     All\'egaten 55, NO-5007 Bergen, Norway
    \label{BERGEN}}
\titlefoot{Dipartimento di Fisica, Universit\`a di Bologna and INFN,
     Via Irnerio 46, IT-40126 Bologna, Italy
    \label{BOLOGNA}}
\titlefoot{Centro Brasileiro de Pesquisas F\'{\i}sicas, rua Xavier Sigaud 150,
     BR-22290 Rio de Janeiro, Brazil \\
     \indent~~and Depto. de F\'{\i}sica, Pont. Univ. Cat\'olica,
     C.P. 38071 BR-22453 Rio de Janeiro, Brazil \\
     \indent~~and Inst. de F\'{\i}sica, Univ. Estadual do Rio de Janeiro,
     rua S\~{a}o Francisco Xavier 524, Rio de Janeiro, Brazil
    \label{BRASIL}}
\titlefoot{Comenius University, Faculty of Mathematics and Physics,
     Mlynska Dolina, SK-84215 Bratislava, Slovakia
    \label{BRATISLAVA}}
\titlefoot{Coll\`ege de France, Lab. de Physique Corpusculaire, IN2P3-CNRS,
     FR-75231 Paris Cedex 05, France
    \label{CDF}}
\titlefoot{CERN, CH-1211 Geneva 23, Switzerland
    \label{CERN}}
\titlefoot{Institut de Recherches Subatomiques, IN2P3 - CNRS/ULP - BP20,
     FR-67037 Strasbourg Cedex, France
    \label{CRN}}
\titlefoot{Now at DESY-Zeuthen, Platanenallee 6, D-15735 Zeuthen, Germany
    \label{DESY}}
\titlefoot{Institute of Nuclear Physics, N.C.S.R. Demokritos,
     P.O. Box 60228, GR-15310 Athens, Greece
    \label{DEMOKRITOS}}
\titlefoot{FZU, Inst. of Phys. of the C.A.S. High Energy Physics Division,
     Na Slovance 2, CZ-180 40, Praha 8, Czech Republic
    \label{FZU}}
\titlefoot{Dipartimento di Fisica, Universit\`a di Genova and INFN,
     Via Dodecaneso 33, IT-16146 Genova, Italy
    \label{GENOVA}}
\titlefoot{Institut des Sciences Nucl\'eaires, IN2P3-CNRS, Universit\'e
     de Grenoble 1, FR-38026 Grenoble Cedex, France
    \label{GRENOBLE}}
\titlefoot{Helsinki Institute of Physics, HIP,
     P.O. Box 9, FI-00014 Helsinki, Finland
    \label{HELSINKI}}
\titlefoot{Joint Institute for Nuclear Research, Dubna, Head Post
     Office, P.O. Box 79, RU-101 000 Moscow, Russian Federation
    \label{JINR}}
\titlefoot{Institut f\"ur Experimentelle Kernphysik,
     Universit\"at Karlsruhe, Postfach 6980, DE-76128 Karlsruhe,
     Germany
    \label{KARLSRUHE}}
\titlefoot{Institute of Nuclear Physics and University of Mining and Metalurgy,
     Ul. Kawiory 26a, PL-30055 Krakow, Poland
    \label{KRAKOW}}
\titlefoot{Universit\'e de Paris-Sud, Lab. de l'Acc\'el\'erateur
     Lin\'eaire, IN2P3-CNRS, B\^{a}t. 200, FR-91405 Orsay Cedex, France
    \label{LAL}}
\titlefoot{School of Physics and Chemistry, University of Lancaster,
     Lancaster LA1 4YB, UK
    \label{LANCASTER}}
\titlefoot{LIP, IST, FCUL - Av. Elias Garcia, 14-$1^{o}$,
     PT-1000 Lisboa Codex, Portugal
    \label{LIP}}
\titlefoot{Department of Physics, University of Liverpool, P.O.
     Box 147, Liverpool L69 3BX, UK
    \label{LIVERPOOL}}
\titlefoot{LPNHE, IN2P3-CNRS, Univ.~Paris VI et VII, Tour 33 (RdC),
     4 place Jussieu, FR-75252 Paris Cedex 05, France
    \label{LPNHE}}
\titlefoot{Department of Physics, University of Lund,
     S\"olvegatan 14, SE-223 63 Lund, Sweden
    \label{LUND}}
\titlefoot{Universit\'e Claude Bernard de Lyon, IPNL, IN2P3-CNRS,
     FR-69622 Villeurbanne Cedex, France
    \label{LYON}}
\titlefoot{Univ. d'Aix - Marseille II - CPP, IN2P3-CNRS,
     FR-13288 Marseille Cedex 09, France
    \label{MARSEILLE}}
\titlefoot{Dipartimento di Fisica, Universit\`a di Milano and INFN,
     Via Celoria 16, IT-20133 Milan, Italy
    \label{MILANO}}
\titlefoot{Universit\`a degli Studi di Milano - Bicocca,
     Via Emanuelli 15, IT-20126 Milan, Italy
    \label{MILANO2}}
\titlefoot{Niels Bohr Institute, Blegdamsvej 17,
     DK-2100 Copenhagen {\O}, Denmark
    \label{NBI}}
\titlefoot{IPNP of MFF, Charles Univ., Areal MFF,
     V Holesovickach 2, CZ-180 00, Praha 8, Czech Republic
    \label{NC}}
\titlefoot{NIKHEF, Postbus 41882, NL-1009 DB
     Amsterdam, The Netherlands
    \label{NIKHEF}}
\titlefoot{National Technical University, Physics Department,
     Zografou Campus, GR-15773 Athens, Greece
    \label{NTU-ATHENS}}
\titlefoot{Physics Department, University of Oslo, Blindern,
     NO-1000 Oslo 3, Norway
    \label{OSLO}}
\titlefoot{Dpto. Fisica, Univ. Oviedo, Avda. Calvo Sotelo
     s/n, ES-33007 Oviedo, Spain
    \label{OVIEDO}}
\titlefoot{Department of Physics, University of Oxford,
     Keble Road, Oxford OX1 3RH, UK
    \label{OXFORD}}
\titlefoot{Dipartimento di Fisica, Universit\`a di Padova and
     INFN, Via Marzolo 8, IT-35131 Padua, Italy
    \label{PADOVA}}
\titlefoot{Rutherford Appleton Laboratory, Chilton, Didcot
     OX11 OQX, UK
    \label{RAL}}
\titlefoot{Dipartimento di Fisica, Universit\`a di Roma II and
     INFN, Tor Vergata, IT-00173 Rome, Italy
    \label{ROMA2}}
\titlefoot{Dipartimento di Fisica, Universit\`a di Roma III and
     INFN, Via della Vasca Navale 84, IT-00146 Rome, Italy
    \label{ROMA3}}
\titlefoot{DAPNIA/Service de Physique des Particules,
     CEA-Saclay, FR-91191 Gif-sur-Yvette Cedex, France
    \label{SACLAY}}
\titlefoot{Instituto de Fisica de Cantabria (CSIC-UC), Avda.
     los Castros s/n, ES-39006 Santander, Spain
    \label{SANTANDER}}
\titlefoot{Dipartimento di Fisica, Universit\`a degli Studi di Roma
     La Sapienza, Piazzale Aldo Moro 2, IT-00185 Rome, Italy
    \label{SAPIENZA}}
\titlefoot{Inst. for High Energy Physics, Serpukov
     P.O. Box 35, Protvino, (Moscow Region), Russian Federation
    \label{SERPUKHOV}}
\titlefoot{J. Stefan Institute, Jamova 39, SI-1000 Ljubljana, Slovenia
     and Laboratory for Astroparticle Physics,\\
     \indent~~Nova Gorica Polytechnic, Kostanjeviska 16a, SI-5000 Nova Gorica, Slovenia, \\
     \indent~~and Department of Physics, University of Ljubljana,
     SI-1000 Ljubljana, Slovenia
    \label{SLOVENIJA}}
\titlefoot{Fysikum, Stockholm University,
     Box 6730, SE-113 85 Stockholm, Sweden
    \label{STOCKHOLM}}
\titlefoot{Dipartimento di Fisica Sperimentale, Universit\`a di
     Torino and INFN, Via P. Giuria 1, IT-10125 Turin, Italy
    \label{TORINO}}
\titlefoot{Dipartimento di Fisica, Universit\`a di Trieste and
     INFN, Via A. Valerio 2, IT-34127 Trieste, Italy \\
     \indent~~and Istituto di Fisica, Universit\`a di Udine,
     IT-33100 Udine, Italy
    \label{TU}}
\titlefoot{Univ. Federal do Rio de Janeiro, C.P. 68528
     Cidade Univ., Ilha do Fund\~ao
     BR-21945-970 Rio de Janeiro, Brazil
    \label{UFRJ}}
\titlefoot{Department of Radiation Sciences, University of
     Uppsala, P.O. Box 535, SE-751 21 Uppsala, Sweden
    \label{UPPSALA}}
\titlefoot{IFIC, Valencia-CSIC, and D.F.A.M.N., U. de Valencia,
     Avda. Dr. Moliner 50, ES-46100 Burjassot (Valencia), Spain
    \label{VALENCIA}}
\titlefoot{Institut f\"ur Hochenergiephysik, \"Osterr. Akad.
     d. Wissensch., Nikolsdorfergasse 18, AT-1050 Vienna, Austria
    \label{VIENNA}}
\titlefoot{Inst. Nuclear Studies and University of Warsaw, Ul.
     Hoza 69, PL-00681 Warsaw, Poland
    \label{WARSZAWA}}
\titlefoot{Fachbereich Physik, University of Wuppertal, Postfach
     100 127, DE-42097 Wuppertal, Germany
    \label{WUPPERTAL}}
\addtolength{\textheight}{-10mm}
\addtolength{\footskip}{5mm}
\clearpage
\headsep 30.0pt
\end{titlepage}
%
\pagenumbering{arabic} 
\setcounter{footnote}{0} %
\large
%
\section{Introduction}
   The study of the process $e^+e^-\to e^+e^-\mu^+\mu^-$ 
provides a good way to test QED up to the fourth order of 
$\alpha$.
The photon structure can be studied by measuring photon structure functions 
which can be extracted in the so-called ``single tagged'' mode (Fig.~1), 
where one of the scattered electrons\footnote[1]{Throughout the paper 
the term ``electron'' will be used for the tagged electron/positron}
is detected (``tagged'') in an electromagnetic calorimeter
while the other scattered electron goes undetected (``untagged'').
This process can also
be used as a reference one for studies of the hadronic structure function
of the photon, providing a basis for a better understanding of the detector 
performance and for checking the analysis procedure. \\
\indent
   Previous measurements of muon pair production in both the single tagged mode
and the double tagged mode (where the scattered electron and positron are both 
detected) have shown good agreement with QED predictions
\cite{CELLO,TPC,PLUTO,JADE,MARKJ,CELLO1,MARKII,HRS,AMY,OP1,Poz,OP2,L3,OP3},
with one exception~\cite{AMY} where
an excess of data events was observed in the double tag mode. 

This study, based on the data collected by the DELPHI collaboration 
at LEP1 at centre-of-mass energies from 89.4 to 93~GeV, complements those results.
It improves on previous DELPHI measurements of the leptonic photon structure function
$F_2^\gamma$ \cite{Poz} by including all the LEP1 statistics and increasing the
$Q^2$ coverage by an order of magnitude. This paper also presents 
results of studies of the azimuthal correlations, which are used to 
extract the ratios of the structure functions $F_A^\gamma/F_2^\gamma$ 
and $F_B^\gamma/F_2^\gamma$. \\

\section{Event kinematics}
In the single tagged mode, where the tagged and untagged electrons 
are scattered with polar angles $\theta_{tag}$ and $\theta_{untag}$ 
and energies $E_{tag}$ and $E_{untag}$ respectively,
and the probe and target photons have four-momenta $q=(x_{tag}E_{beam},\vec{q})$
and $p=(x_{untag}E_{beam},\vec{p})$, 
the cross section of the reaction $e^+e^-\to e^+e^-X$ 
is given by \cite{Budnev}:
\begin{equation}
 \frac{d^3\sigma}{dxdQ^2dx_{untag}} = \frac{dn(x_{untag})}{dx_{untag}} \times
   \frac{d^2\sigma(e\gamma\to eX)}{dxdQ^2}.
\end{equation}
These two factors, the flux of target photons and the $e\gamma$ cross
section respectively, are given by:
\begin{equation}
\frac{dn(x_{untag})}{dx_{untag}} = \frac{\alpha}{\pi x_{untag}}
\left\{ \left(1+(1-x_{untag})^2\right) \ln\left(\frac{2E_{beam}(1-x_{untag})}
{m_ex_{untag}}\sin\frac{\theta_{untag}^{max}}{2}\right)-1+x_{untag} \right\},
\end{equation}
\begin{equation}
 \frac{d^2\sigma(e\gamma\to eX)}{dxdQ^2} = \frac{2\pi\alpha^2}{xQ^4} 
\left\{ \left(1+(1-y)^2\right) F_2^\gamma(x,Q^2,P^2)-y^2F_L^\gamma(x,Q^2,P^2)
 \right\}.
\end{equation}
Here $F_2^\gamma$ and $F_L^\gamma$ are structure functions of the photon,
$\alpha$ is the QED coupling constant,
$Q^2 = -q^2 \simeq 4E_{tag}E_{beam}\sin^2(\theta_{tag}/2)$ 
is the squared 4-momentum transfer,
$P^2=-p^2$ is the virtuality of the target photon, 
and $x$ and $y$ are the Bjorken variables
\begin{equation}
x = \frac{Q^2}{2q\cdot p} = \frac{Q^2}{W_{\gamma\gamma}^2+Q^2+P^2},
~~~~~~~~y = \frac{p\cdot q}{p\cdot k} \simeq \frac{Q^2}{sxx_{untag}} = 
1 - \frac{E_{tag}}{E_{beam}}\cos^2\frac{\theta_{tag}}{2}
\end{equation}
where $W_{\gamma\gamma}^2=(q+p)^2$ is the invariant mass of the $\gamma\gamma$
(or $\mu^+\mu^-$) system, 
$k$ is the initial four-vector of the tagged electron, and $s=4E_{beam}^2$.

The structure function $F_2^\gamma$ can be extracted
from the dependence of the cross section on $x$ and $Q^2$.
But $F_L^\gamma$ is small and is weighted by the small factor $y^2$, 
making its direct measurement impractical. 

However, additional structure functions can be studied 
by looking at azimuthal correlations of the final state particles.
The differential cross section of the process can be written as~\cite{LEP2}
\begin{eqnarray}
\frac{d^4\sigma(e\gamma\to e\mu^+\mu^-)}{dxdyd\cos\theta^*d\chi/4\pi} & = &
\frac{2\pi\alpha^2}{Q^2}\cdot\frac{1+(1-y)^2}{xy} \times \nonumber\\
 & & \left\{ (2x\tilde F_T+
\epsilon(y)\tilde F_L)-\rho(y)\tilde F_A\cos\chi+\frac{1}{2}
\epsilon(y)\tilde F_B\cos2\chi \right\}, 
\end{eqnarray}
where $\chi$ is the azimuthal angle, defined in the $\gamma\gamma^*$ 
centre-of-mass frame as the angle between the planes formed by 
the photon axis and the muon and the scattered electron respectively
(Fig.~2), 
and $\theta^*$ is the angle 
between the muon and the photon axis. The functions $\rho(y)$ and $\epsilon(y)$
are given by $\rho(y)=(2-y)\sqrt{1-y}/
(1+(1-y)^2)$ and $\epsilon(y)=2(1-y)/(1+(1-y)^2)$ \cite{Zerwas}
and can be taken equal to 1 in the accessible kinematical region.
The differential structure functions 
$\tilde F_T$, $\tilde F_L$, $\tilde F_A$, and $\tilde F_B$ 
give the corresponding standard structure functions 
$F_T$, $F_L$, $F_A$, and $F_B$ 
after integrating appropriately over $\cos\theta^*$ (see section 7.2) 
taking into account that $F_A$ is antisymmetric in $\cos\theta^*$ \cite{AZIM}.
The cross section can then be written as
\begin{eqnarray}
 \frac{d^3\sigma(e\gamma\to e\mu^+\mu^-)}{dxdyd\chi/2\pi} & \simeq & 
\frac{2\pi\alpha^2}{Q^2}\cdot\frac{1+(1-y)^2}{xy} \times \nonumber \\
 & & F_2^\gamma\left( 1-(F_A^\gamma/F_2^\gamma)\cos\chi+
\frac{1}{2}(F_B^\gamma/F_2^\gamma)\cos2\chi \right). 
\end{eqnarray}
The structure functions $F_i^\gamma$ are combinations of transition 
amplitudes for the different helicity states of the photons. The structure
function $F_B^\gamma$ is related to the interference term between the two
transverse helicity states of the photons. It is identical to $F_L^\gamma$, 
which is related to the longitudinal polarization of the virtual photon, 
in leading order and for massless muons. \\ 

\section{DELPHI detector}
The DELPHI detector has been described in detail elsewhere \cite{DELPHI,PERF}. 
In this analysis, the scattered electron was tagged using
\begin{itemize}
\item the Small Angle Tagger (SAT), the main luminosity monitor during
1991-93, covering polar angles from 2.5$^\circ$ to 8$^\circ$ 
(172$^\circ$ to 177.5$^\circ$); it was made of
alternating layers of lead sheets (0.9 mm thick) and plastic scintillator
fibres (1 mm in diameter), aligned parallel to the beam;
\item the Small angle TIle Calorimeter (STIC), the main luminosity 
monitor since 1994, covering polar angles from 
1.7$^\circ$ to 10.3$^\circ$ (169.7$^\circ$ to 178.3$^\circ$);
the STIC is a sampling calorimeter 
with 49 sandwiches of 3.4 mm steel-laminated lead plates and 3 mm 
thick scintillator tiles giving a total thickness of $\sim$27 radiation lengths;
\item the Forward ElectroMagnetic Calorimeter (FEMC) covering from 10$^\circ$
to 36.5$^\circ$ (143.5$^\circ$ to 170$^\circ$) in polar angle, 
consisting of two 5 m diameter
disks containing a total of 9064 lead glass blocks.
\end{itemize}
The energy resolution of the tagging calorimeters was around 5\% in SAT
and FEMC and 3\% in STIC for an incident electron energy of 45 GeV.\\
\indent
   For muon identification, DELPHI contained barrel and forward
muon detectors, each consisting of at least 4 layers of drift chambers.
The muon chambers covered 78\% of the solid angle.\\
\indent
   Combining the information from the tracking detectors, the relative
momentum resolution $\sigma_p/p$ varied from 0.001$\times p$ to
0.01$\times p$ ($p$ in GeV/$c$), depending on the polar angle of the
charged particle. 

\section{Monte Carlo simulation}
   Two event generators were used in order to simulate 
the signal process $e^+e^-\to e^+e^-\mu^+\mu^-$:
BDKRC \cite{BDKRC} which includes only the multiperipheral 
diagram (Fig.1) together with QED radiative corrections, and 
DIAG36 \cite{DIAG36} which lacks the QED radiative corrections 
but includes also the bremsstrahlung, annihilation and conversion diagrams. 
DIAG36 was used  to check the role of these additional diagrams. 

   Several generators were used to estimate the backgrounds to
the process studied: 
BDKRC \cite{BDKRC} was used to simulate $e^+e^-\to e^+e^-\tau^+\tau^-$,
TWOGAM \cite{TWOGAM} to simulate hadron production in two-photon collisions, 
DYMU3 \cite{DYMU3} for the $e^+e^-\to \mu^+\mu^-(\gamma)$ process, 
and KORALZ \cite{KORALZ} for $e^+e^-\to \tau^+\tau^-(\gamma)$. 

   The generated events were passed through the full simulation 
of the DELPHI detector and reconstructed using the same program as for the 
data. 

\section{Event selection and correction}
   Events were selected as single tagged dimuon candidates if the following
requirements were met.
\begin{itemize}
\item There was a cluster in one of the electromagnetic calorimeters
with an energy deposition greater than 0.6$\times E_{beam}$ (hereafter
called the tagged electron). If the cluster lay 
within the polar angle range 20$^\circ$ - 160$^\circ$, it 
was linked to a detected charged particle. 
\item There were exactly two additional particles with
opposite charges and polar angles between 20$^\circ$ and 160$^\circ$.
The relative errors on their momenta were less than 1.
Their impact parameters with respect to the average interaction
point were below 4 cm in the transverse plane and 10 cm along the beam. 
Their track lengths seen in the tracking detectors were at least 30 cm. 
Their momenta were above 0.5 GeV/$c$ and 2.5 GeV/$c$ 
and the sum of their momenta was below 30 GeV/$c$. 
\item At least one of the additional particles with a momentum
greater than 2.5 GeV/$c$ was identified as a muon by the DELPHI standard 
muon tagging algorithm \cite{PERF}.
\item The invariant mass of the two additional particles was 
above 1.7 GeV/$c^2$. 
This requirement  reduced the contribution from diagrams other than the
multiperipheral one  to below 0.25\% for the low $Q^2$ and 2\% for the high
$Q^2$ sample according to the DIAG36 generator, 
and avoided possible problems  with the soft part of the spectrum due to
trigger or muon tagging inefficiency.
\item Finally, double-tagged events were rejected 
by requiring there to be no energy deposit exceeding 
0.3$\times E_{beam}$ in the detector arm (defined as 
$\theta=0^\circ-90^\circ$ and $\theta=90^\circ-180^\circ$) opposite
that containing the tagged electron.
\end{itemize}
\hspace*{10pt}
Using the high redundancy of the trigger~\cite{PERF}, the trigger inefficiency
was found to be negligible for these events.\\
\indent
   In order to improve the measurements of the tagged electron parameters 
(energy and angles), the following procedures were used.
\begin{enumerate}
\item To avoid edge effects, the tagged electron was required to lie in the 
polar angle range $3^\circ<\theta<7.6^\circ$ ($172.4^\circ<\theta<177^\circ$)
for the SAT, $2.5^\circ<\theta<9^\circ$ ($171^\circ<\theta<177.5^\circ$) 
for the STIC, or $11^\circ<\theta<35^\circ$ ($145^\circ<\theta<169^\circ$)
for the FEMC.
\item To improve the $\theta$ measurements in the SAT, which had a limited
granularity, the radial position of the cluster was corrected using
the function found from the comparison of the experimental 
radial distribution for Bhabha events with the theoretical one based on 
a $1/\theta^3$ cross section dependence (Fig.~\ref{Radius}).
This improved the $Q^2$ resolution from 6.0\% to 2.9\%.
\item To improve the $\theta$ measurements in the SAT and STIC, 
their alignments were checked using Bhabha event samples. The detector
on the electron side had a mask in front of it to better define the acceptance
at low $\theta$. From the number of Bhabha events as a function
of the electron azimuthal angle $\phi_1$, it was possible to find the 
displacement of the mask relative to the beam line. 
The alignment on the opposite side
was checked by looking at the difference of the measured polar angles 
$\theta_{tag}-\theta_{untag}$ of the scattered electron and positron 
as a function of the positron azimuthal angle $\phi_2$ (Fig.~\ref{Angle}).
The dependencies observed were used to correct the measured polar angles.
The errors of the fitted parameters were taken as uncertainties of the 
procedure, contributing 0.5\% uncertainty on low values of $Q^2$.
\item A more accurate value of the tagged electron energy $E_{tag}$ 
was calculated from the requirements of energy and longitudinal 
momentum conservation in the event:
\begin{equation}
E_{tag} = \frac{P_{\mu\mu}\cos\theta_{\mu\mu}+(2E_{beam}-E_{\mu\mu})
   \cos\theta_{untag}}{\cos\theta_{untag}-\cos\theta_{tag}},
\end{equation}
where $P_{\mu\mu}$, $E_{\mu\mu}$ and $\theta_{\mu\mu}$ are the momentum,
energy and polar angle of the muon system, 
and $\theta_{untag}$ is the polar angle of the untagged electron, 
assumed to be 0 or $\pi$. 
The improvement due to this method can be seen in Fig.~\ref{Ecor}, 
obtained from simulation, where the difference between the reconstructed
and true (generated) tag energy $E_{tag} - E_{tag}^{gen}$ is shown as a 
function of the tag angle $\theta_{tag}$ using both the direct measurement 
of $E_{tag}$ and this method.
\end{enumerate}
 
\section{Background }
   The following sources of background to the $\mu^+\mu^-$ event samples
were considered:
\begin{itemize}
\item $e^+e^-\to e^+e^-\tau^+\tau^-$ with a $\tau$ decay product identified 
as a muon. The background
from this process was found to be (1.2$\pm$0.2)\% 
for the SAT and STIC tagged
samples and (5.7$\pm$1.1)\% for the FEMC, where the errors quoted are
statistical.
\item $e^+e^-\to \tau^+\tau^-(\gamma)$ with a hard radiated photon or a 
$\tau$ decay product faking a tagged electron. This background was found to be
negligible for the SAT and STIC samples, and (8.9$\pm$1.9)\% for the FEMC,
after taking into account the on-peak versus off-peak luminosity distribution
of the data.
\item $e^+e^-\to \mu^+\mu^-(\gamma)$ with the radiated photon  faking a
tagged electron. This  was found to be negligible due to the 30~GeV cut on the
sum of the muon momenta.
\item $e^+e^-\to e^+e^-\pi^+\pi^-$ with a pion misidentified as a muon. 
The ratio of the cross sections for pion pair and muon pair production in 
two-photon interactions falls to (1-5)\% if the invariant mass of the
produced pair is above 2.0 GeV/$c^2$ \cite{pipi}. With the muon 
identification criteria described above, the probability to misidentify
a pion as a muon was below 1.5\% (depending on the pion momentum), so
this background was also negligible for all samples.
\item other $e^+e^-\to e^+e^-+hadrons$ processes. These were also 
found to be negligible for all event samples.
\item untagged $e^+e^-\to e^+e^-\mu^+\mu^-$ in coincidence with an
off-momentum electron faking a tagged electron. The off-momentum
electrons are beam electrons that have scattered off residual gas molecules
inside the beam pipe. Using a method similar to the one described in 
\cite{offmom}, this background was estimated from Z$^0\to \mu\mu$ events
in coincidence with a similar off-momentum electron, multiplying by the
ratio of the dimuon production cross sections from untagged two-photon
interactions and from Z$^0$ decays, and was also found to be negligible.\\
\end{itemize}

\section{Results}
   The numbers of selected data events after background subtraction are
compared with the predictions of the signal Monte Carlo simulations in
Table~1. The $Q^2$ ranges shown are calculated given the angular
coverage of the detectors and the cut on the tag energy, and the average 
values $<Q^2>$ are taken from the data. Figs.~\ref{SAT} -- \ref{FEMC} present
the distributions of a standard set of observables for events tagged by the 
SAT, STIC and FEMC respectively.

\begin{table}[hbtp]\centering
\begin{tabular}{|c|c|c|c|}\hline
Tagging detector           & SAT        & STIC        & FEMC     \\ 
$Q^2$ range (GeV$^2/c^4$)  & 3.4$-$36.6 & 2.4$-$51.2 & 45.9$-$752.8 \\
$<Q^2>^{*)}$ (GeV$^2/c^4$) & 13.0 & 12.1 & 120.0$^{**)}$ \\ \hline
data                    & 1357$\pm$37 & 2875$\pm$54 & 239$\pm$18 \\ 
BDKRC simulation      & 1362$\pm$14 & 2884$\pm$22 & 250$\pm$6 \\ 
DIAG36 simulation     & 1298$\pm$25 & 2785$\pm$55 & 236$\pm$13 \\ \hline
\multicolumn{4}{l} {$^{*)}$ After requiring $E_{tag}>0.75\times E_{beam}$ 
(see text)} \\
\multicolumn{4}{l} {$^{**)}$ For events with $\theta_{tag}<25^\circ~
(\theta_{tag}>155^\circ)$ (see text)} \\
\end{tabular}
\vskip 3mm
\caption{Numbers of selected events after background subtraction. }
\end{table}
\indent
   Table~1 and Figs.~\ref{SAT} - \ref{FEMC} show that
the BDKRC and DIAG36 generators produce similar kinematical
distributions, but DIAG36 gives somewhat lower numbers of selected events. 
In the kinematical region under study, the contribution of the additional
diagrams in DIAG36 was found to be very small (see section 5).
This difference (if real) should therefore be attributed to the effect of
radiative corrections.
The BDKRC generator was therefore used for the structure 
function studies below.

\subsection{Extraction of $F_2^\gamma$}
   To extract $F_2^\gamma$, the experimental $x$ distribution was
divided by the Monte Carlo distribution weighted by the factor
$\alpha/F_2^\gamma(x,Q^2)$, where $F_2^\gamma(x,Q^2)$ can be obtained 
from a simulated event sample using either a generator producing events
according to a given $F_2$ or the photon flux approach described,
for example, in \cite{L3} and briefly outlined below. \\
\indent
   It follows from Eqs.~(1-3) that, neglecting the small contribution
from $y^2$ terms:
\begin{equation}
F_2^\gamma(x,Q^2,P^2) = \frac{d^2\sigma}{dxdQ^2} /{\cal W}(x,Q^2),
\end{equation}
where 
the weight ${\cal W}(x,Q^2)$ is given by
\begin{equation}
{\cal W}(x,Q^2) = \frac{4\pi\alpha^2}{xQ^4}
\int\limits_{x_{untag}^{min}}^{x_{untag}^{max}}\frac{dn(x_{untag})}{dx_{untag}}
(1-y)~dx_{untag}.
\end{equation}
   To calculate the integration limits, the 
fractional energy of the target photon is extracted from the expressions for
$x$ and $W_{\gamma\gamma}$:
\begin{equation}
x_{untag} = \frac{2Q^2/sx-2Q^2/s+\cos\Theta+\cos\theta_{untag}-x_{tag}(1+\cos\Theta)}
{\cos\Theta+\cos\theta_{untag}+x_{tag}(1-\cos\Theta)},
\end{equation}
where 
\begin{equation}
\cos\Theta = \sin\theta_{tag}\sin\theta_{untag}\cos(\Delta\phi) -
\cos\theta_{tag}\cos\theta_{untag},
\end{equation}
and $\Delta\phi$ is the azimuthal angle between the scattered
$e^+$ and $e^-$. In the single tag approximation, 
$\theta_{untag}\simeq 0$ so that (10) becomes:
\begin{equation}
x_{untag} = \frac{2Q^2/sx-2Q^2/s-\cos\theta_{tag}+1-x_{tag}(1-\cos\theta_{tag})}
{-\cos\theta_{tag}+1+x_{tag}(1+\cos\theta_{tag})}.
\end{equation}
   The maximum and minimum $x_{untag}$ values correspond to the
minimum and maximum $x_{tag}$ values, 
and these result from the tagging conditions:
\begin{equation}
x_{tag}^{max} = \min\left\{1-\frac{E_{tag}^{min}}{E_{beam}},1-\frac{Q^2}{s\sin^2
(\theta_{tag}^{max}/2)}\right\},~~~~
x_{tag}^{min} = \max\left\{\frac{W_{\gamma\gamma}^2}{s},
1-\frac{Q^2}{s\sin^2(\theta_{tag}^{min}/2)}\right\}
\end{equation}
where $E_{tag}^{min}$ is the lower cut on the tag energy and $\theta_{tag}^{min} 
(\theta_{tag}^{max})$ is the lower (upper) angular acceptance of the 
tagging device. $E_{tag}^{min}$ was increased from $0.6\times E_{beam}$ to 
$0.75\times E_{beam}$ in order to keep the $y^2$ contribution small. \\
\indent
   Fig.~\ref{F2gen} shows the $F_2^\gamma(x)$ values obtained by both
methods for a simulated event sample with STIC tagging conditions, 
demonstrating that they give similar results. 

A fit to the QED prediction~\cite{Budnev,Berger}
\begin{equation}
F_2^\gamma = \frac{\alpha}{\pi}x \left\{\left(x^2+(1-x)^2\right)
\ln\frac{W^2_{\gamma\gamma}}{m^2_\mu+P^2x(1-x)} - 1 +8x(1-x)
-\frac{P^2x(1-x)}{m^2_\mu+P^2x(1-x)} \right\},
\end{equation}
where terms of order $m^2_\mu/Q^2$ are neglected
gives values of the effective average target photon virtuality $P^2$ 
of 0.022$\pm$0.007 and 0.026$\pm$0.006~GeV$^2$
for the first and second methods respectively, the
errors quoted being statistical.
For the SAT tagged events the first method, which was chosen 
for the further analysis, gives $P^2$=0.032$\pm$0.007~GeV$^2$,  
demonstrating the need to take the target photon
virtuality into account in studies of $F_2^\gamma$. \\
\indent
   The extracted structure function $<F_2^\gamma(x,Q^2)>$, 
transformed to $F_2^\gamma(x,<Q^2>)$ using the ratio 
$F_2^\gamma(x,<Q^2>)/<F_2^\gamma(x,Q^2)>$ predicted by QED,
is shown in Table~2 and Fig.~\ref{F2exp}, 
which present the weighted combination of the 
SAT and STIC results with $<Q^2>=12.5$~GeV$^2/c^4$ and the FEMC 
result with $<Q^2>=120$~GeV$^2/c^4$. The FEMC sample included only
events with $\theta_{tag}$ below 25$^\circ$ (above 155$^\circ$) in
order to exclude the region with large background contamination 
(Fig.~\ref{FEMC}b), and the contribution from diagrams other than the
multiperipheral one predicted by the BDKRC generator was subtracted.
The structure function values have been corrected to the centres 
of the $x$ bins by
multiplying the measured average values of $F_2^\gamma$ 
for each $x$ bin by the ratio of the value of $F_2^\gamma$ 
in the centre of the bin to the its average value over the bin
predicted by QED.
Systematic errors due to the resolutions in $Q^2$ and $x$ have been evaluated 
in simulation by varying these variables according to their 
resolutions and checking the effect on $F_2^\gamma$. The role of the observed 
discrepancy between the data and simulation in some $\theta_{tag}$ intervals 
(Fig.~\ref{Radius}) was checked by weighting the contributions 
of events in those intervals according to their $\theta_{tag}$ values when 
producing the $x$ distribution. The largest contribution to the systematic
error comes from the $Q^2$ resolution. \\
\indent
   Fits to the QED prediction (14) give $P^2=0.025\pm0.005$ and 
$0.073\pm0.056$~GeV$^2$ for the samples with low and high $Q^2$ respectively, 
in good agreement with the Monte Carlo prediction. \\

\begin{table}[hbtp]\centering
\begin{tabular}{|c|c|c|c|c|c|c|c|c|c|}\hline
$x$ & $<$0.1 & 0.1-0.2 & 0.2-0.3 & 0.3-0.4 & 0.4-0.5 & 0.5-0.6 & 0.6-0.7 
& 0.7-0.8 & $>$0.8 \\ \hline
$F_2^\gamma/\alpha$ & 0.106 & 0.273 & 0.426 & 0.515 & 0.573 & 0.645 & 0.743
& 0.942 & 1.152 \\ 
stat. error & $\pm$0.008 & $\pm$0.012 & $\pm$0.017 & $\pm$0.021 & $\pm$0.024 
& $\pm$0.029 & $\pm$0.038 & $\pm$0.060 & $\pm$0.112 \\ 
syst. error & $\pm$0.023 & $\pm$0.012 & $\pm$0.012 & $\pm$0.012 & $\pm$0.004 
& $\pm$0.003 & $\pm$0.021 & $\pm$0.053 & $\pm$0.094 \\ \hline\hline
\end{tabular}
\vskip 3mm

\begin{tabular}{|c|c|c|c|c|c|}\hline
$x$                 & $<$0.2 & 0.2-0.4 & 0.4-0.6 & 0.6-0.8 & $>$0.8\\ \hline
$F_2^\gamma/\alpha$ & 0.387  & 0.464   & 0.673   & 0.984   & 1.508 \\ 
stat. error & $\pm$0.214 & $\pm$0.133 & $\pm$0.138 & $\pm$0.162 & $\pm$0.231 \\
syst. error & $\pm$0.015 & $\pm$0.051 & $\pm$0.049 & $\pm$0.026 & $\pm$0.044 \\
\hline
\end{tabular}
\vskip 3mm
\caption{The measured structure function $F_2^\gamma$ for 
$<Q^2>=12.5$ (upper table) and 120~GeV$^2/c^4$ (lower table). }
\end{table}

\subsection{Azimuthal correlations}
   In order to increase the observed azimuthal correlations 
of the final state particles, only events 
with $20^\circ<\theta^*<160^\circ$ have been considered.   
Taking into account the antisymmetry of $F_A^\gamma$ in $\cos\theta^*$, events
with $\cos\theta^*<$0 and $\cos\theta^*>$0 have been combined using the
transformation  $\chi\to\pi-\chi$. \\
\indent
   The selected samples have been corrected for detector acceptance
and efficiency using either bin-by-bin corrections over a 
two-dimensional grid of $\chi$ and $\theta^*$, or a three-dimensional 
unfolding \cite{Agostini} in the space of the variables $\chi$, $\theta^*$ 
and $x$. The corrected distributions (Fig.~\ref{cor}) were fitted 
to the expression:
\begin{equation}
dN/d\chi = C~(1+P_1\cos\chi + P_2\cos2\chi)
\end{equation}
where $P_1$ and $P_2$ are closely related to $F_A^\gamma/F_2^\gamma$ and
$F_B^\gamma/F_2^\gamma$,  c.f. Eq.~(6).
The combined results were obtained by refitting the weighted sums of corrected
distributions for the SAT and STIC samples (Fig.~\ref{cor_comb}).
The parameters determined from the fit are shown in Table~3. \\
\indent
   The systematic effects were estimated using simulated events, varying the
variables $Q^2$, $W_{\gamma\gamma}$, $x$, $\theta^*$ and $\chi$ according 
to their resolution, and adding the resulting variations of the fitted
parameters in quadrature. This gave errors on the fitted parameters of 
about 0.02. The difference between the results obtained with the two 
different correction methods gave an additional systematic error of 
0.02$-$0.06. \\
\indent
   The results obtained were extrapolated to the full $\theta^*$ 
and $W_{\gamma\gamma}$ ranges using the theoretical correction factors 
$C_{P_1}$ and $C_{P_2}$ shown in Table 3, which were obtained as ratios 
of the QED predicted structure functions \cite{AZIM} calculated for 
event samples generated in the $Q^2$ range of 2.4-51.2~GeV$^2$
without and with the selection cuts. The results thus obtained for 
$F_A^\gamma/F_2^\gamma$ and ${1\over2}F_B^\gamma/F_2^\gamma$ are shown in 
Table~3 and Fig.~\ref{corvsx}. They are in agreement with the theoretical
predictions \cite{AZIM} and with the results of 
other LEP experiments \cite{L3,OP3} (note the factor -1/2 difference
of $F_A^\gamma$ with \cite{L3} due to its 
different definition). 

\section{Conclusions}
   Muon pair production in single-tagged $\gamma\gamma$ collisions 
has been studied at $\sqrt{s}\simeq$91~GeV using data 
collected by the DELPHI detector at LEP during the years 1992-95.
Distributions of different event variables for $Q^2$ ranging from
$\sim$2.5 to $\sim$750~GeV$^2/c^4$ are well reproduced 
by a Monte Carlo simulation based on QED. \\
\indent
   The leptonic structure function $F_2^\gamma$ has been measured for
two regions of momentum transfer with average $Q^2$ values of
12.5 and 120~GeV$^2/c^4$.\\
\indent
   Azimuthal correlations of final state particles have also been studied,
giving information on additional structure functions $F_A^\gamma$
and $F_B^\gamma$. The measured ratios $F_A^\gamma/F_2^\gamma$ and
$F_B^\gamma/F_2^\gamma$ are significantly different
from zero and consistent with QED expectations.

\newpage
\subsection*{Acknowledgements}
\vskip 3 mm
 We wish to thank V.~Andreev and Ch.~Carimalo for useful discussions.\\
 We are greatly indebted to our technical 
collaborators, to the members of the CERN-SL Division for the excellent 
performance of the LEP collider, and to the funding agencies for their
support in building and operating the DELPHI detector.\\
We acknowledge in particular the support of \\
Austrian Federal Ministry of Science and Traffics, GZ 616.364/2-III/2a/98, \\
FNRS--FWO, Belgium,  \\
FINEP, CNPq, CAPES, FUJB and FAPERJ, Brazil, \\
Czech Ministry of Industry and Trade, GA CR 202/96/0450 and GA AVCR A1010521,\\
Danish Natural Research Council, \\
Commission of the European Communities (DG XII), \\
Direction des Sciences de la Mati$\grave{\mbox{\rm e}}$re, CEA, France, \\
Bundesministerium f$\ddot{\mbox{\rm u}}$r Bildung, Wissenschaft, Forschung 
und Technologie, Germany,\\
General Secretariat for Research and Technology, Greece, \\
National Science Foundation (NWO) and Foundation for Research on Matter (FOM),
The Netherlands, \\
Norwegian Research Council,  \\
State Committee for Scientific Research, Poland, 2P03B06015, 2P03B1116 and
SPUB/P03/178/98, \\
JNICT--Junta Nacional de Investiga\c{c}\~{a}o Cient\'{\i}fica 
e Tecnol$\acute{\mbox{\rm o}}$gica, Portugal, \\
Vedecka grantova agentura MS SR, Slovakia, Nr. 95/5195/134, \\
Ministry of Science and Technology of the Republic of Slovenia, \\
CICYT, Spain, AEN96--1661 and AEN96-1681,  \\
The Swedish Natural Science Research Council,      \\
Particle Physics and Astronomy Research Council, UK, \\
Department of Energy, USA, DE--FG02--94ER40817. \\

\newpage
\begin{table}[hbtp]\centering
\begin{tabular}{|c|c|c|c|c|}\hline
$x$-interval & \multicolumn{3}{c|}{$P_1$} & $C_{P_1}$ \\ \cline{2-4}
             & SAT & STIC & Combined      &           \\ \hline
$x<0.2$      & ~$0.19\pm0.14\pm0.03$ & ~$0.28\pm0.08\pm0.04$ &
               ~0.25$\pm$0.08 & 0.541 \\ 
$0.2<x<0.4$  & ~$0.22\pm0.09\pm0.03$ & ~$0.20\pm0.06\pm0.02$ &
               ~0.20$\pm$0.05 & 0.701 \\ 
$0.4<x<0.6$  & ~$0.13\pm0.09\pm0.05$ & ~$0.02\pm0.07\pm0.05$ &
               ~0.06$\pm$0.07 & 0.625 \\ 
$x>0.6$      & -$0.41\pm0.10\pm0.07$ & -$0.26\pm0.07\pm0.05$ &
               -0.31$\pm$0.07 & 0.849 \\ 
all $x$      & -$0.03\pm0.05\pm0.03$ & -$0.03\pm0.03\pm0.03$ &
               -0.03$\pm$0.04 & 0.605 \\ \hline \hline
\end{tabular}

\vskip 3mm
\begin{tabular}{|c|c|c|c|c|}\hline
$x$-interval & \multicolumn{3}{c|}{$P_2$} & $C_{P_2}$ \\ \cline{2-4}
             & SAT & STIC & Combined      &           \\ \hline
$x<0.2$      & $0.06\pm0.12\pm0.03$ & -$0.01\pm0.08\pm0.03$ &
               0.01$\pm$0.07 & 0.391 \\ 
$0.2<x<0.4$  & $0.13\pm0.08\pm0.03$ & ~$0.16\pm0.06\pm0.02$ &
               0.15$\pm$0.05 & 0.512 \\ 
$0.4<x<0.6$  & $0.15\pm0.08\pm0.04$ & ~$0.19\pm0.06\pm0.03$ &
               0.17$\pm$0.06 & 0.581 \\ 
$x>0.6$      & $0.20\pm0.09\pm0.06$ & ~$0.30\pm0.06\pm0.04$ &
               0.27$\pm$0.06 & 0.673 \\ 
all $x$      & $0.13\pm0.05\pm0.02$ & ~$0.15\pm0.03\pm0.02$ &
               0.15$\pm$0.03 & 0.570 \\ \hline \hline
\end{tabular}

\vskip 3mm
\begin{tabular}{|c|c|c|}\hline
$x$-interval & $F_A^\gamma/F_2^\gamma$ & ${1\over2} F_B^\gamma/F_2^\gamma$ \\ \hline
$x<0.2$      & ~$0.135\pm0.043$        & $0.004\pm0.027$ \\ 
$0.2<x<0.4$  & ~$0.140\pm0.035$        & $0.077\pm0.026$ \\ 
$0.4<x<0.6$  & ~$0.038\pm0.044$        & $0.099\pm0.035$ \\ 
$x>0.6$      & -$0.263\pm0.059$        & $0.182\pm0.040$ \\ 
all $x$      & -$0.018\pm0.024$        & $0.086\pm0.017$ \\ \hline
\end{tabular}
\vskip 3mm
\caption{Parameters $P_1$ and $P_2$ of the fit to the azimuthal 
angle distributions for the SAT-tagged, STIC-tagged, and combined 
event samples with $Q^2=2.4-51.2$~GeV$^2$.
The first error is statistical and the second is systematic.
$C_{P_1}$ and $C_{P_2}$ are the correction factors 
to extrapolate the parameters to the full $\theta^*$ range (see text).
The values extracted 
for $F_A^\gamma/F_2^\gamma$ and ${1\over2} F_B^\gamma/F_2^\gamma$ 
are shown with statistical and systematic errors added in quadrature.}
\end{table}

\clearpage
\hspace*{-1cm}
\begin{tabular}{ll}
\epsfig{figure=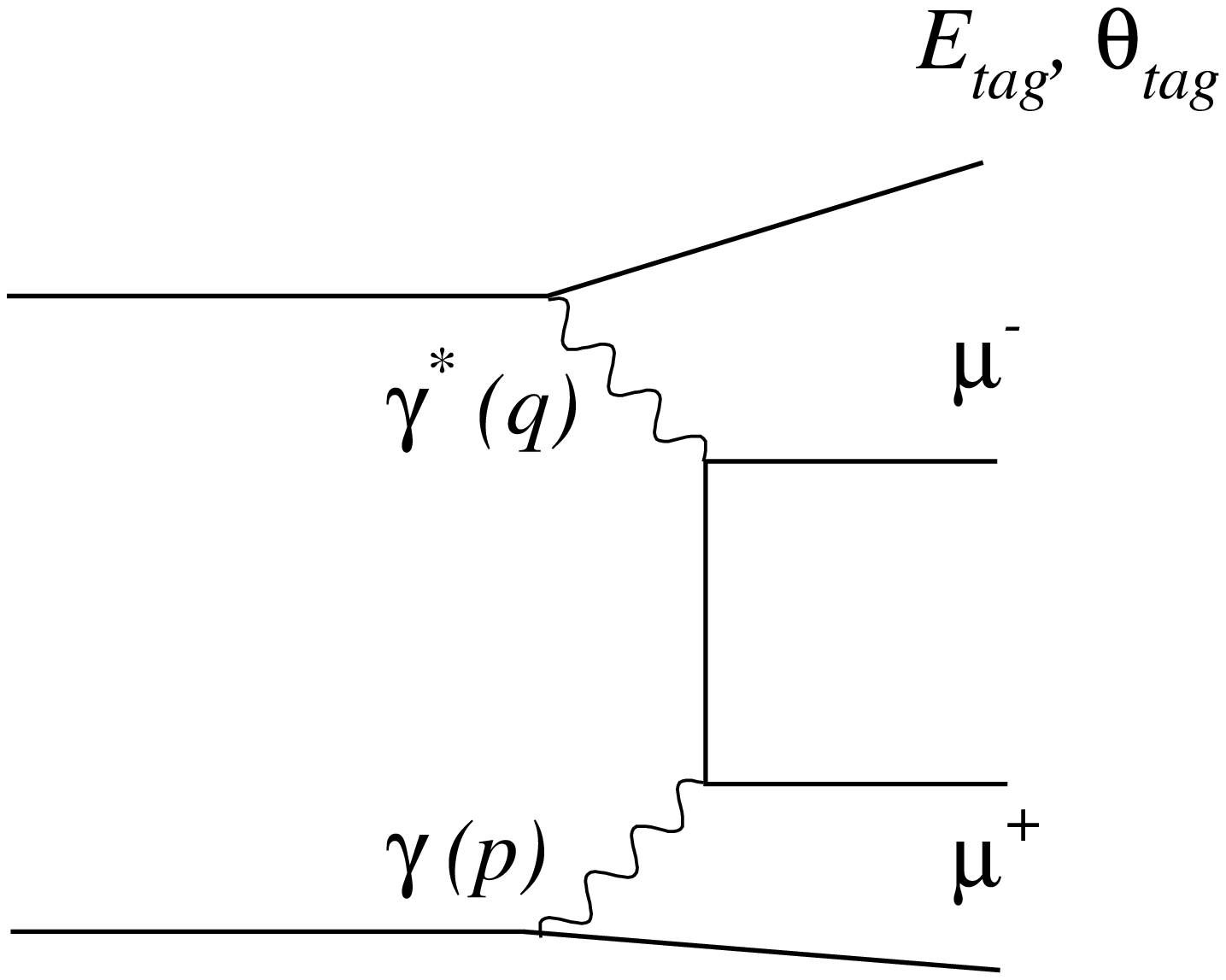,width=9cm} & \hspace*{-10mm}
\epsfig{figure=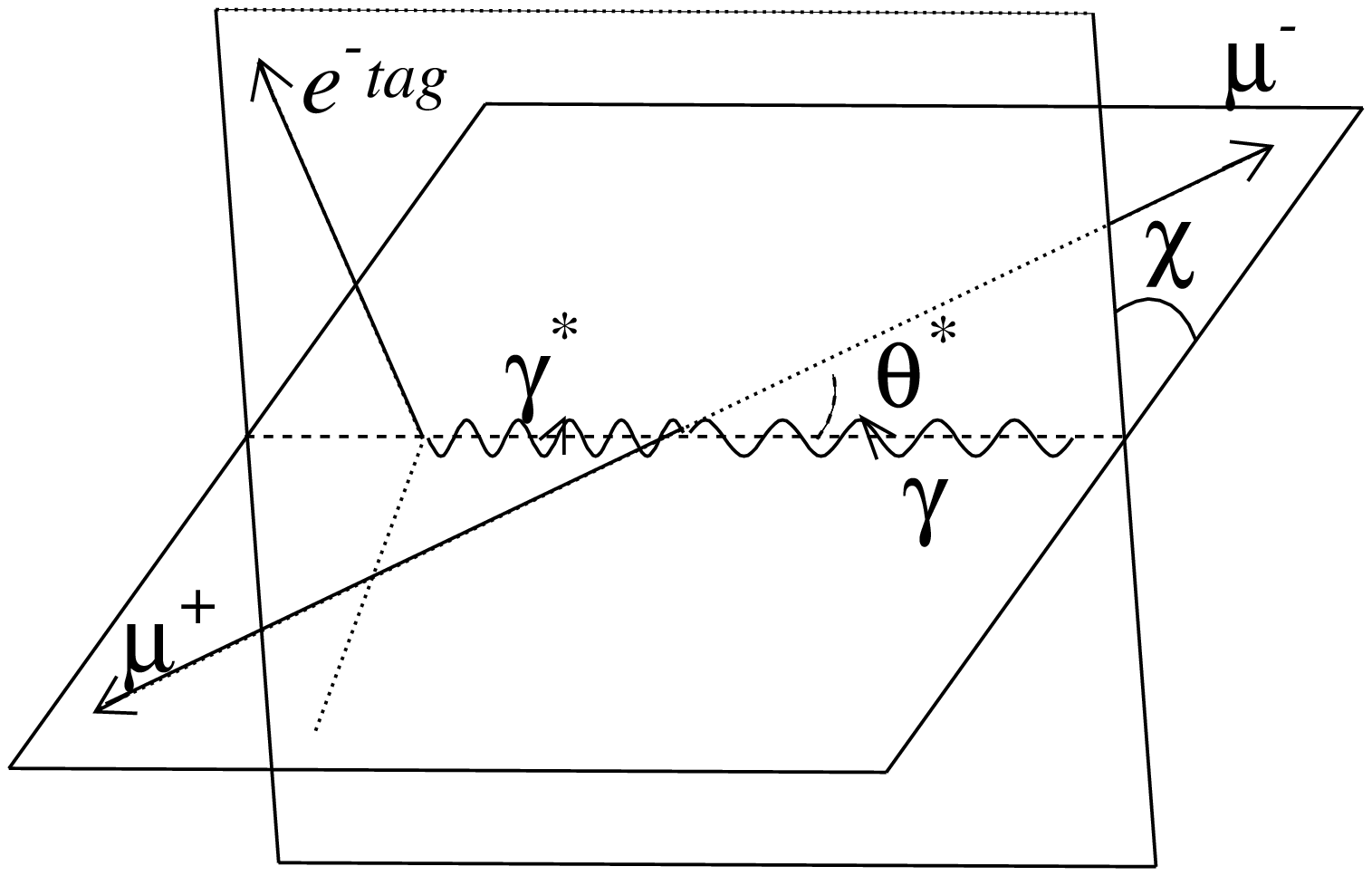,width=9cm} \\
\hspace*{10pt}\parbox[c]{7.5cm}
{ Figure~1: The dominant multiperipheral diagram for the reaction 
$e^+e^-\to e^+e^-\mu^+\mu^-$. $E_{tag}$ and $\theta_{tag}$ are the 
energy and scattering angle of the tagged electron or positron. } &
\hspace{0cm}\parbox[b]{7.cm}
{Figure~2: Definitions of the angles $\chi$ and $\theta^*$ in the 
$\gamma\gamma^*$ centre-of-mass system. }
\end{tabular}
\setcounter{figure}{2}



\begin{figure}[hb]
\begin{center}
{\Large DELPHI}
\mbox{\epsfig{file=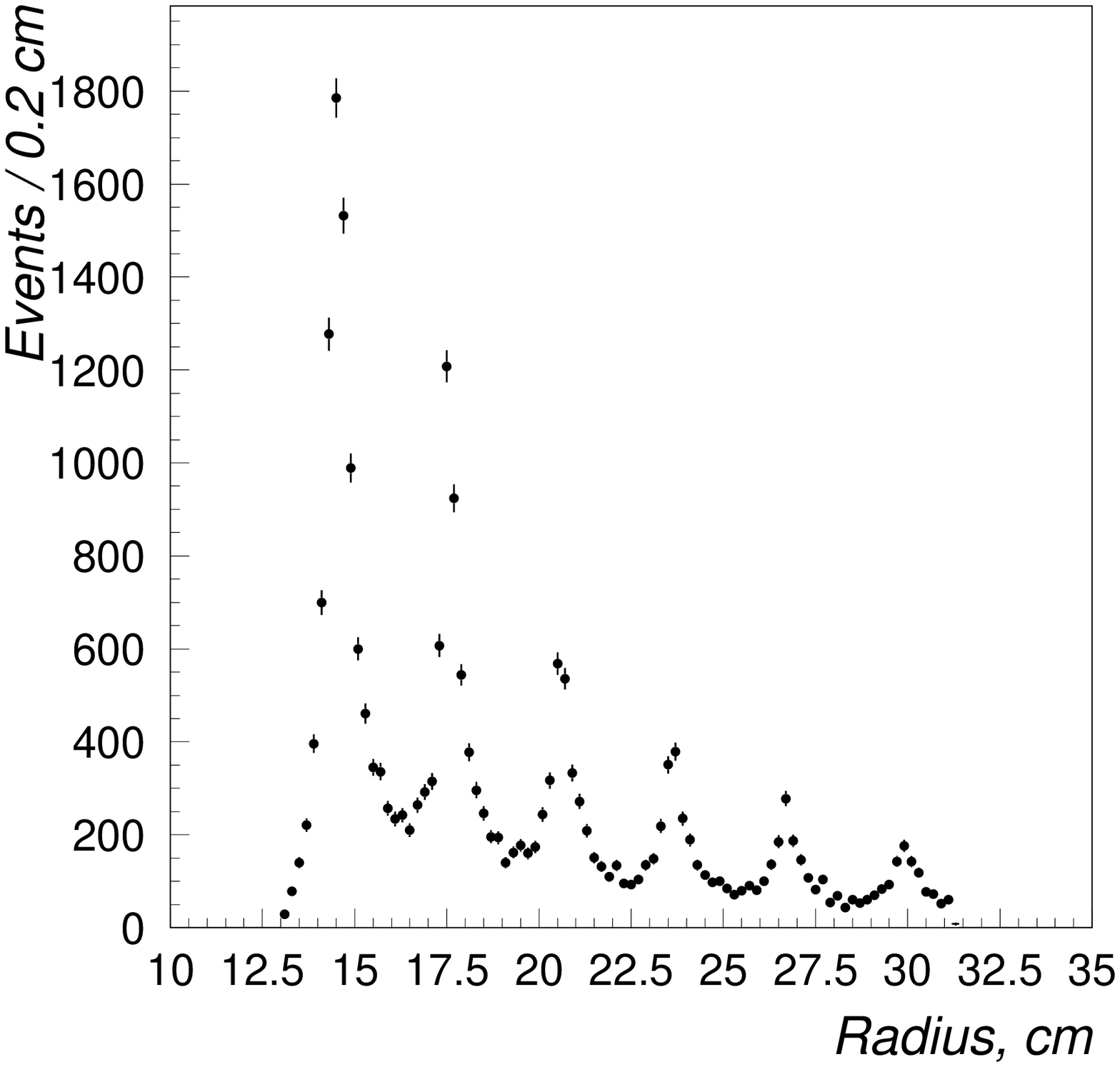,width=8cm }
      \epsfig{file=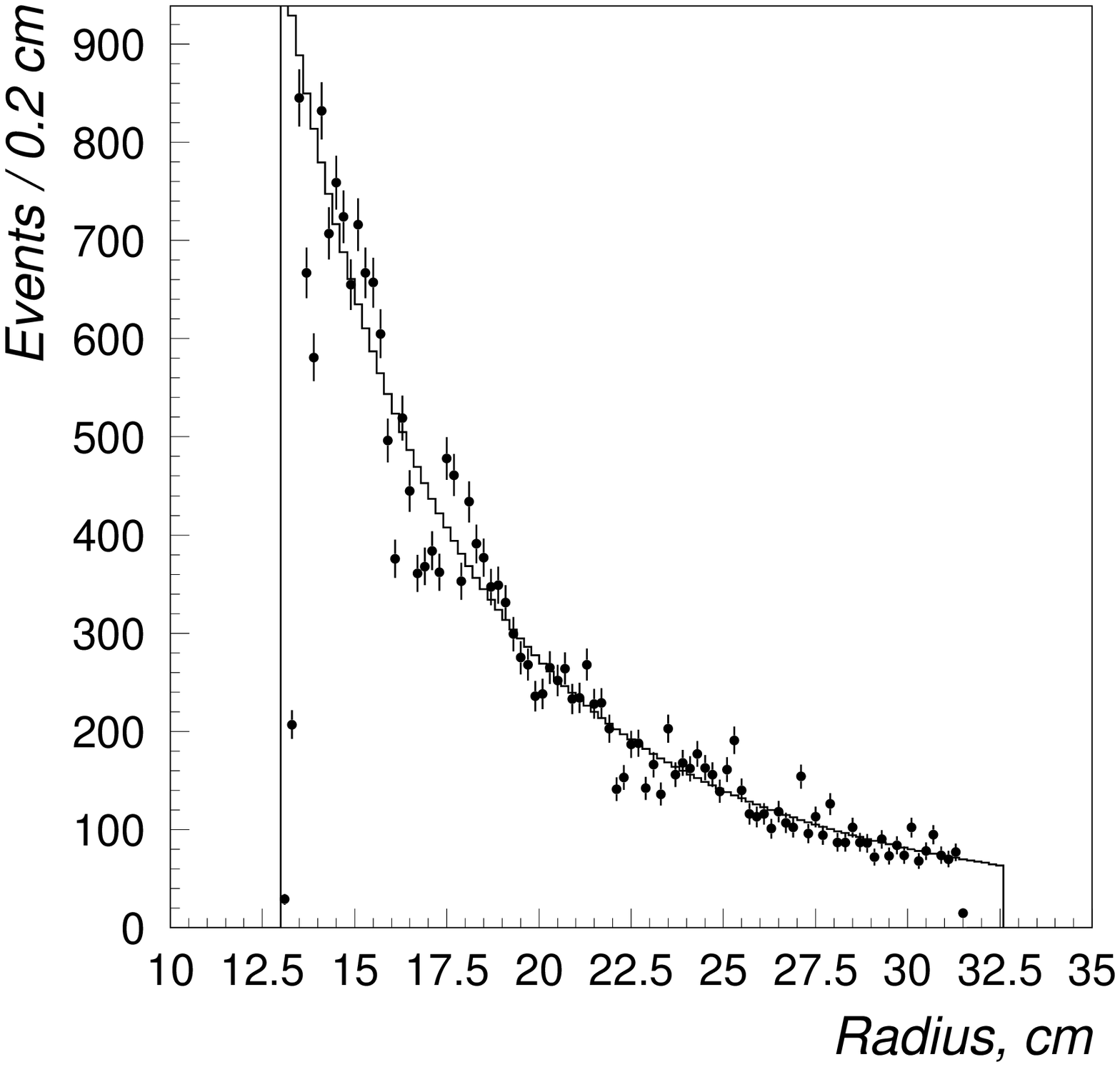,width=8cm }}
\end{center}
\caption{Distributions of the radial positions of Bhabha electrons
in the SAT: 
left - before correcting the radial cluster positions, right - after correction.
The line  shows the result of the fit to the theoretical prediction. }
\label{Radius}
\end{figure}

\clearpage
\begin{figure}[t]
\begin{center}
{\Large DELPHI}
\mbox{\epsfig{file=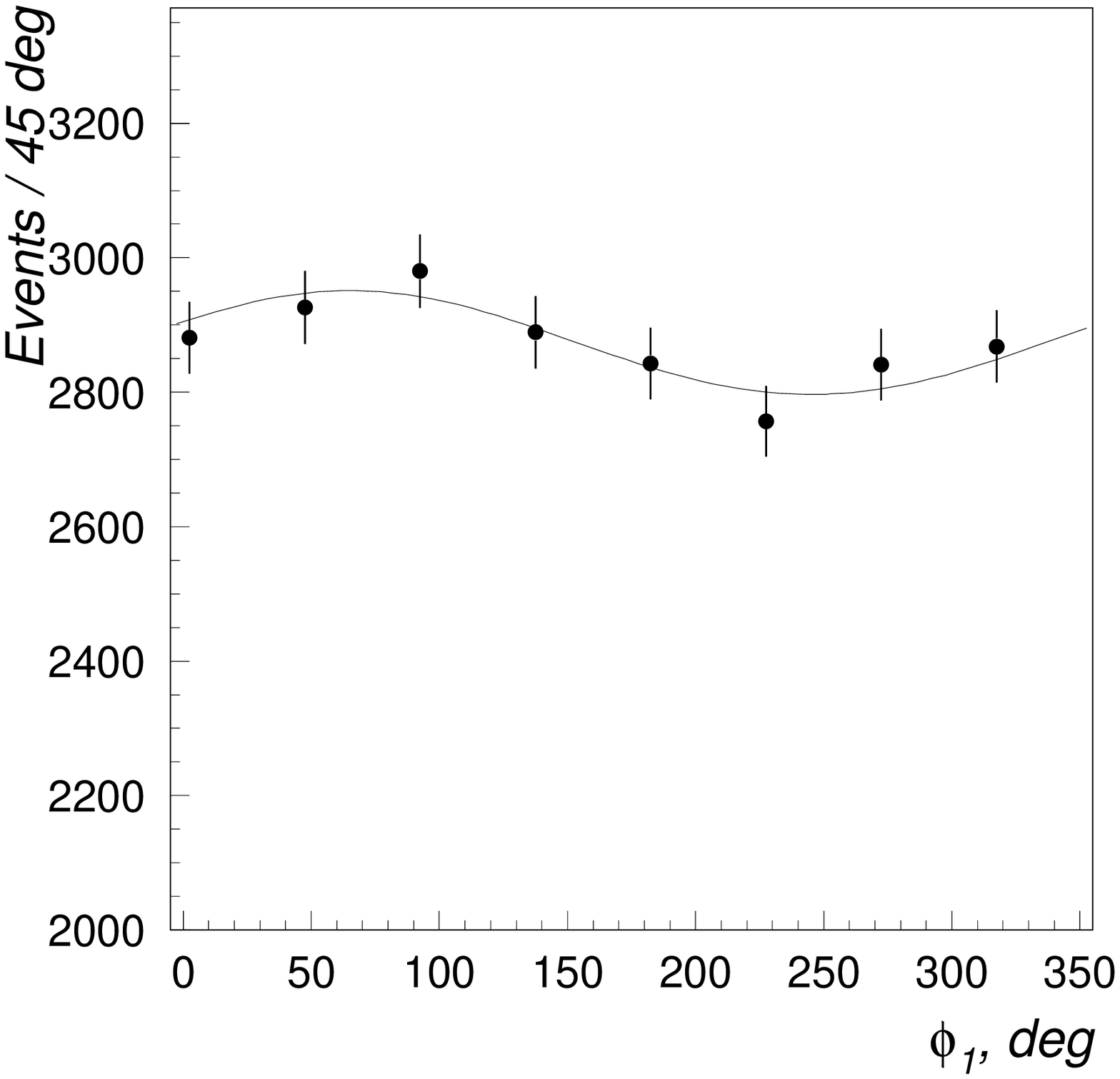,width=8cm }
      \epsfig{file=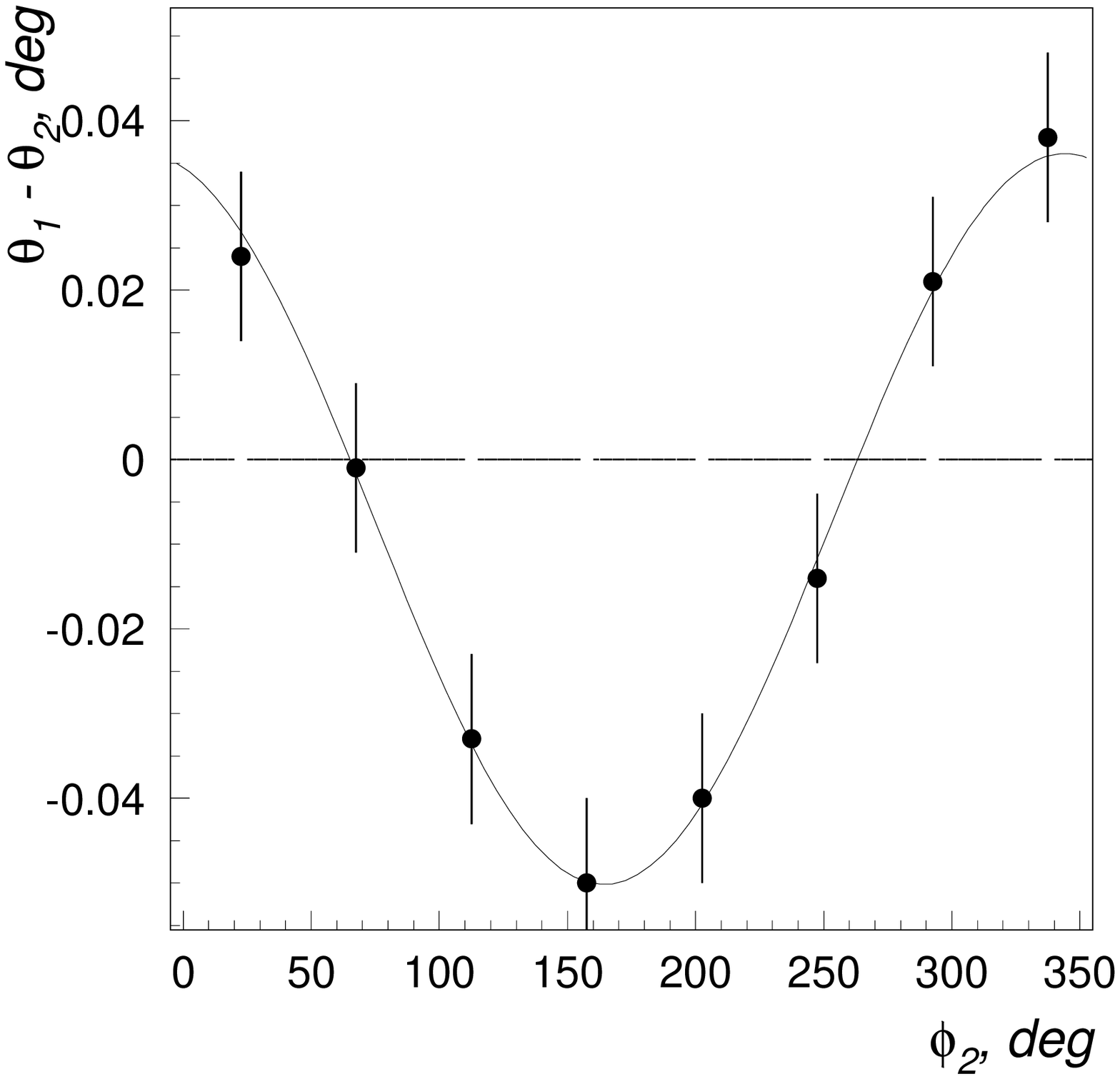,width=8cm }}
\end{center}
\caption{Left - number of detected Bhabha events as a function of the azimuthal
angle for the SAT module with the mask. Right - difference of the polar 
angles of Bhabha electrons measured by the SAT modules as a function of the 
azimuthal angle. The lines show the results of the fits.}
\label{Angle}
\end{figure}

\begin{figure}
\begin{center}
{\Large DELPHI}
\mbox{\epsfig{file=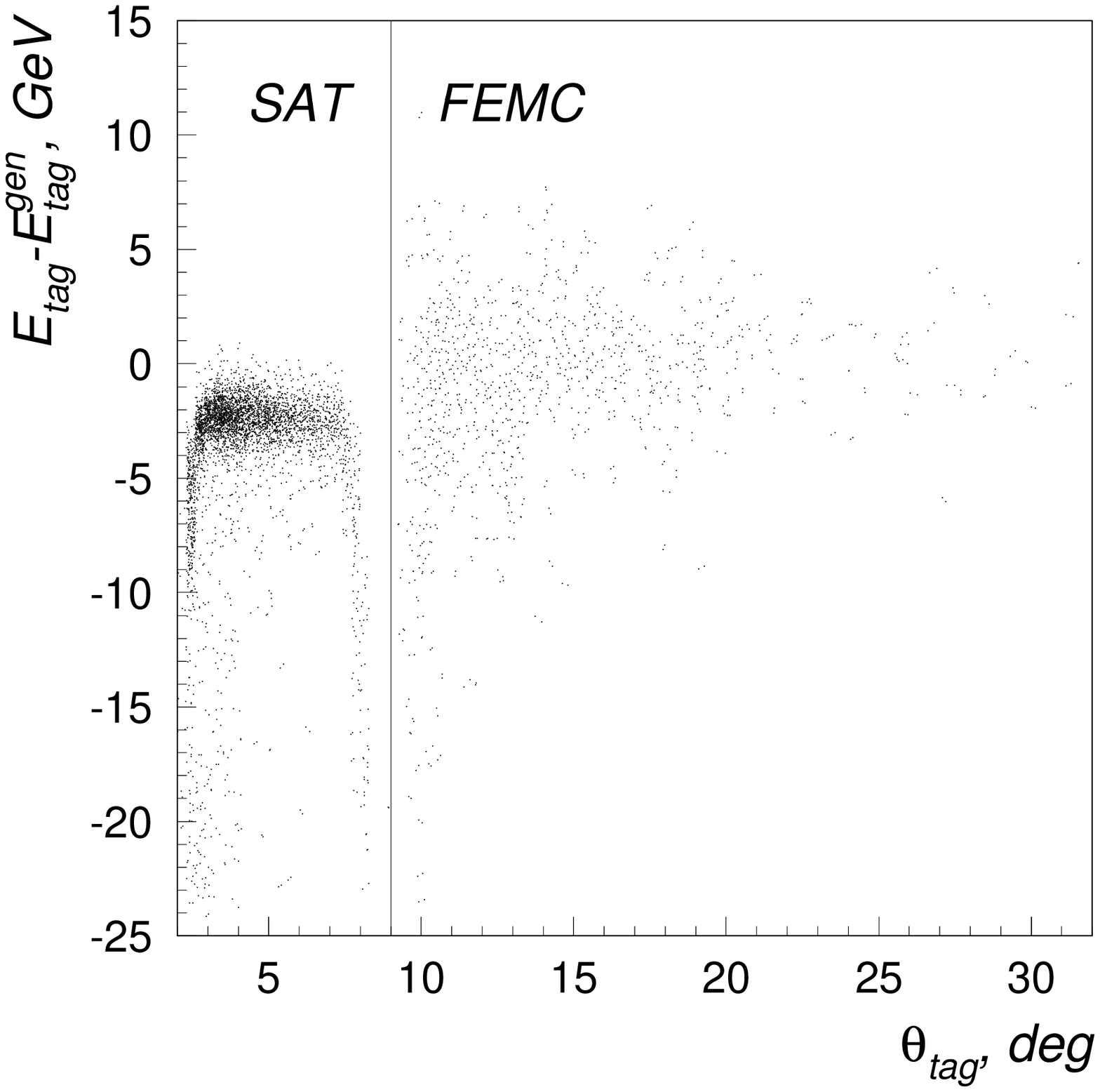,width=8cm }
      \epsfig{file=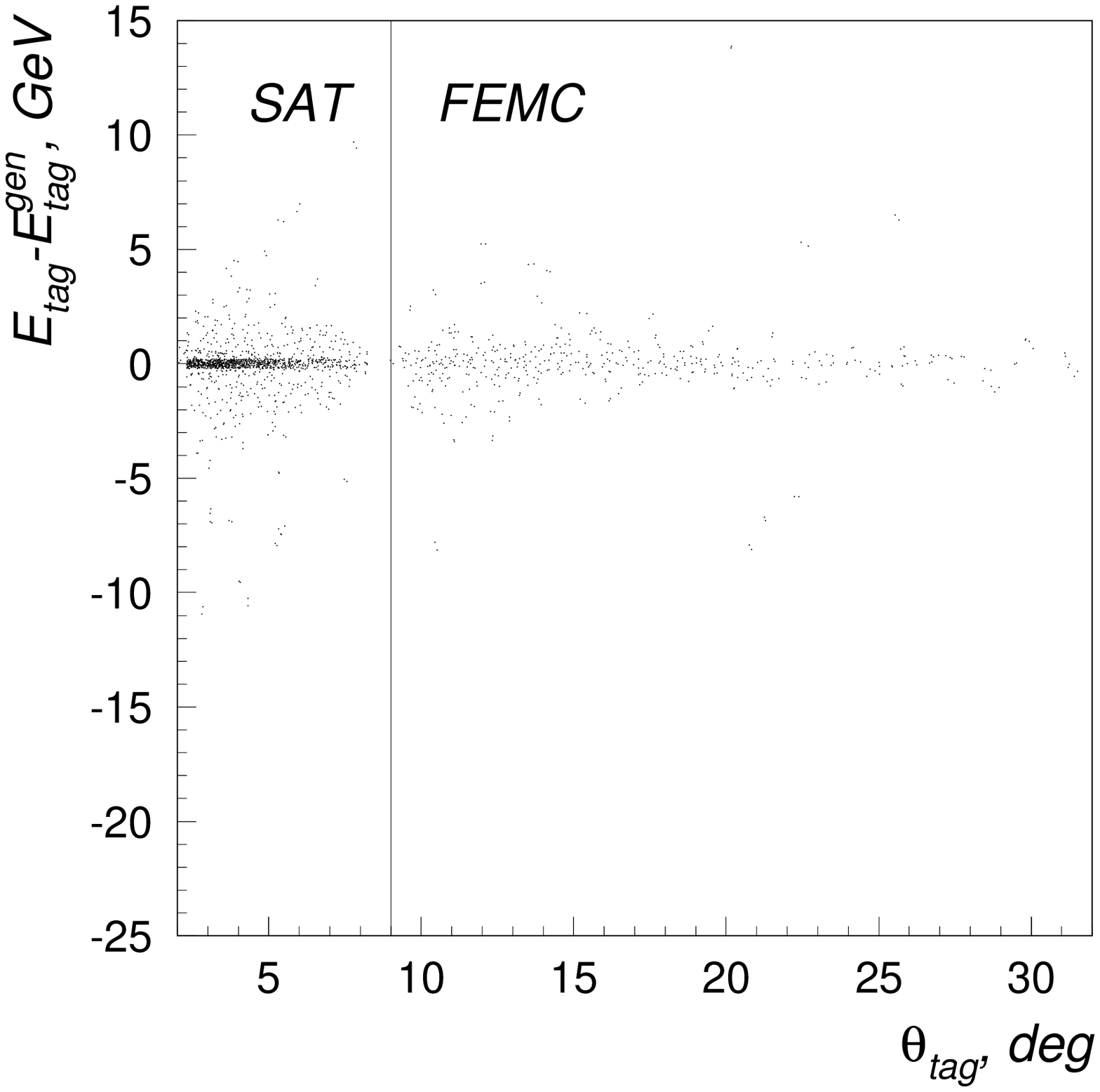,width=8cm }}
\end{center}
\caption{Difference of reconstructed and true (generated) tag energy $E_{tag}$ 
versus the tag angle $\theta_{tag}$ in simulated events: 
left - measured value of $E_{tag}$,
right - $E_{tag}$ value calculated from the kinematics of the event. }
\label{Ecor}
\end{figure}

\begin{figure}
\begin{center}
{\Large DELPHI}
\mbox{\epsfig{file=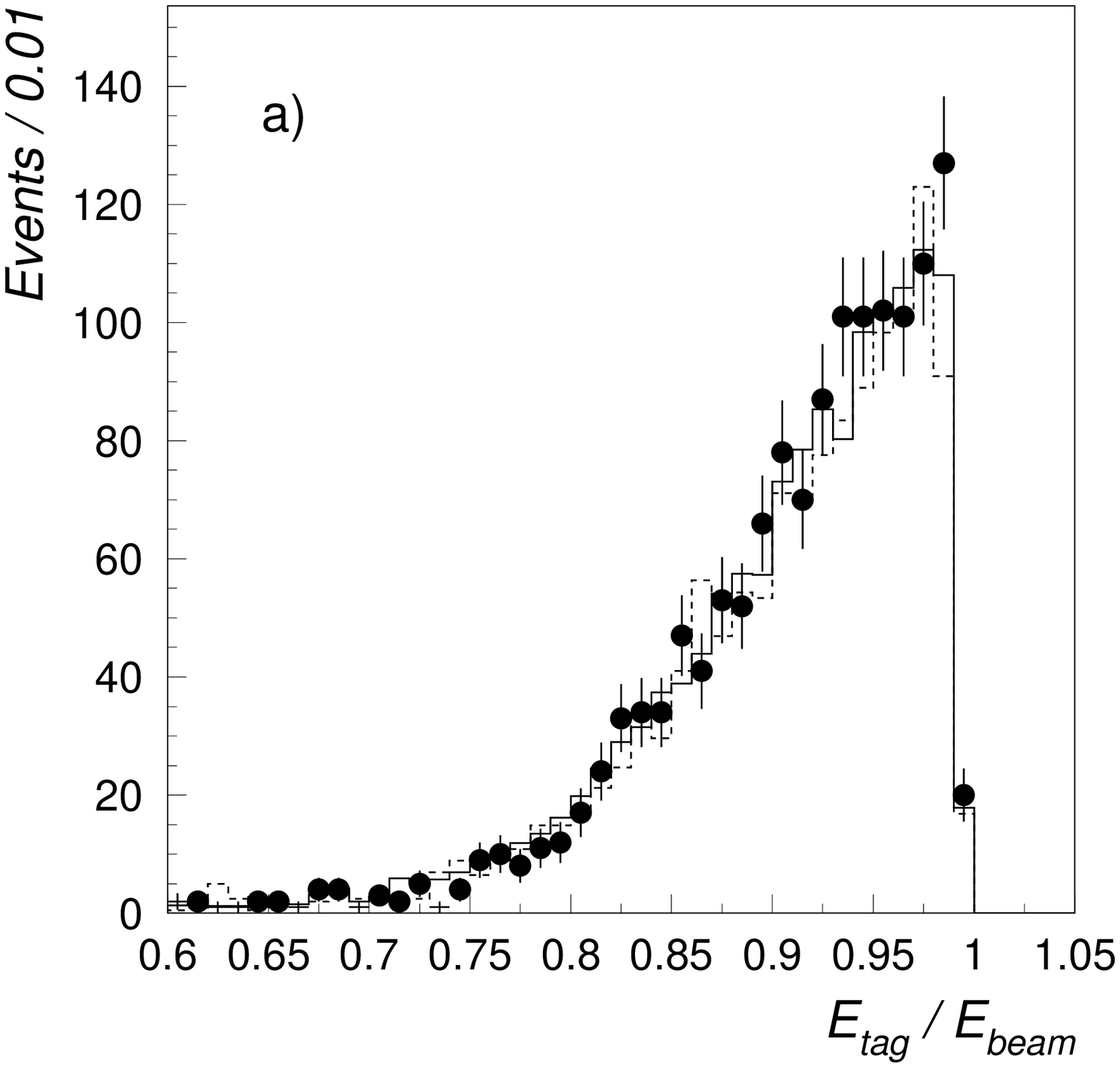,width=7cm }
      \epsfig{file=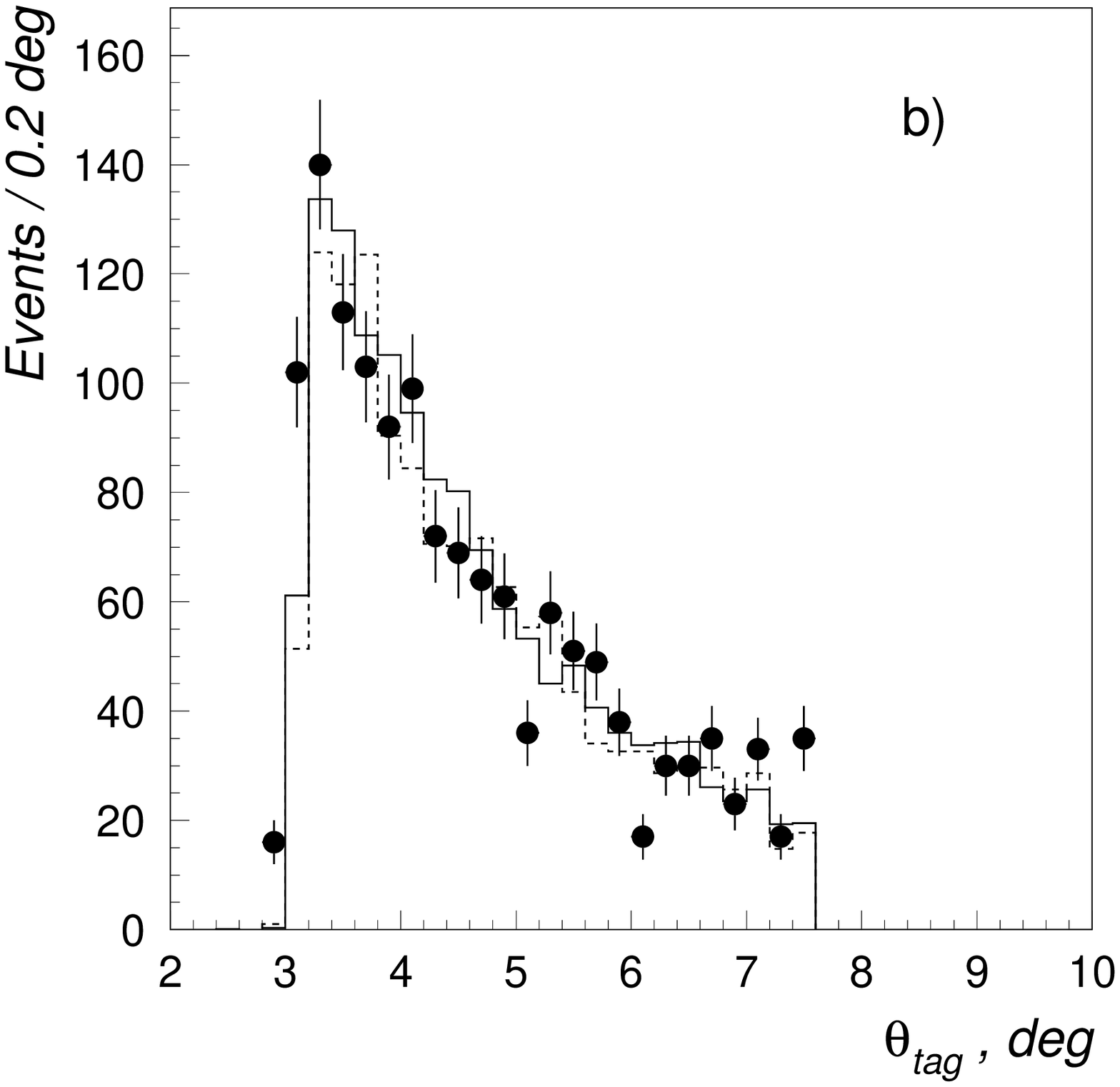,width=7cm }}
\end{center}
\vspace*{-1cm}
\begin{center}
\mbox{\epsfig{file=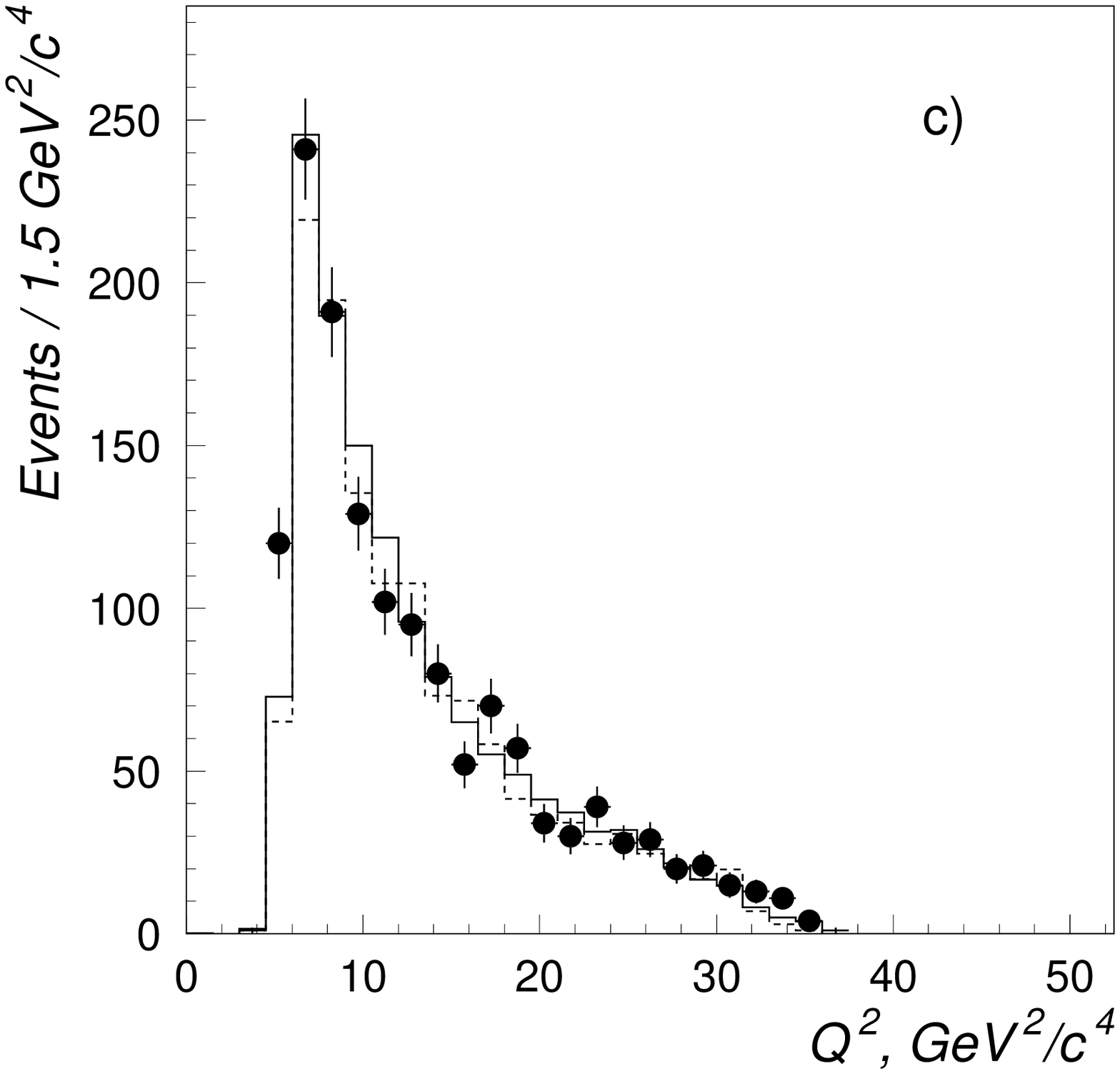,width=7cm }
      \epsfig{file=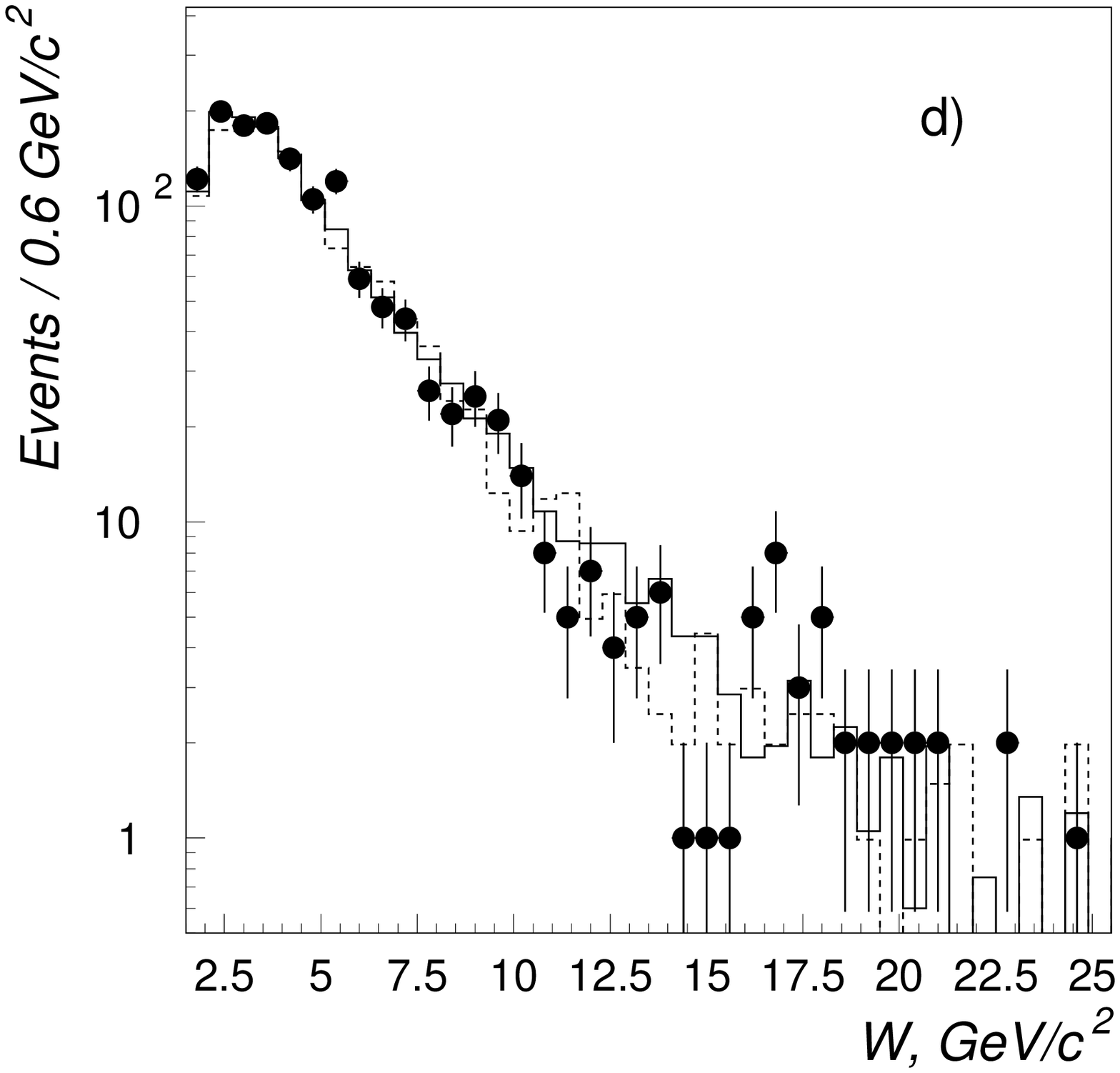,width=7cm }}
\end{center}
\vspace*{-1cm}
\begin{center}
\mbox{\epsfig{file=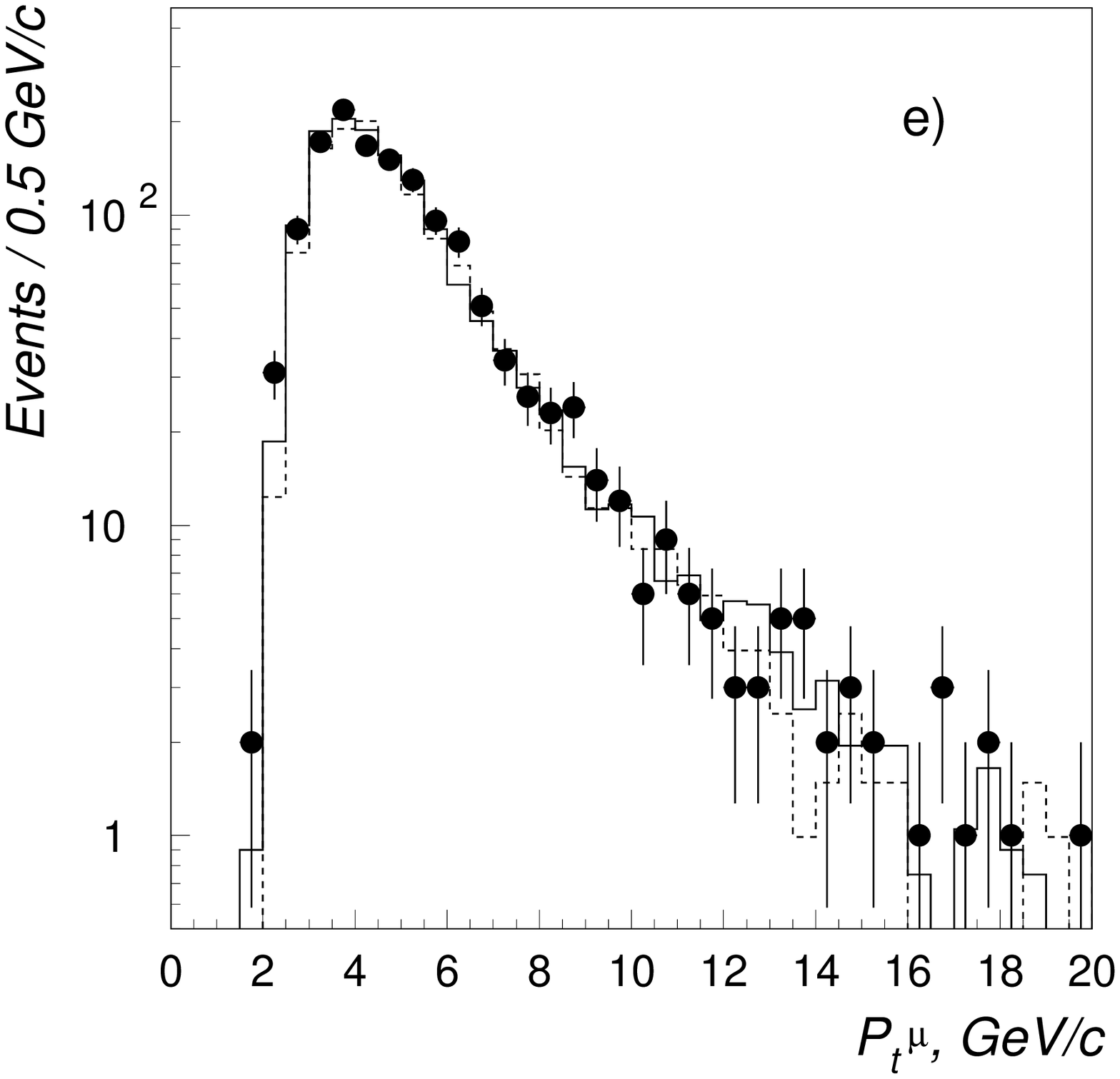,width=7cm }
      \epsfig{file=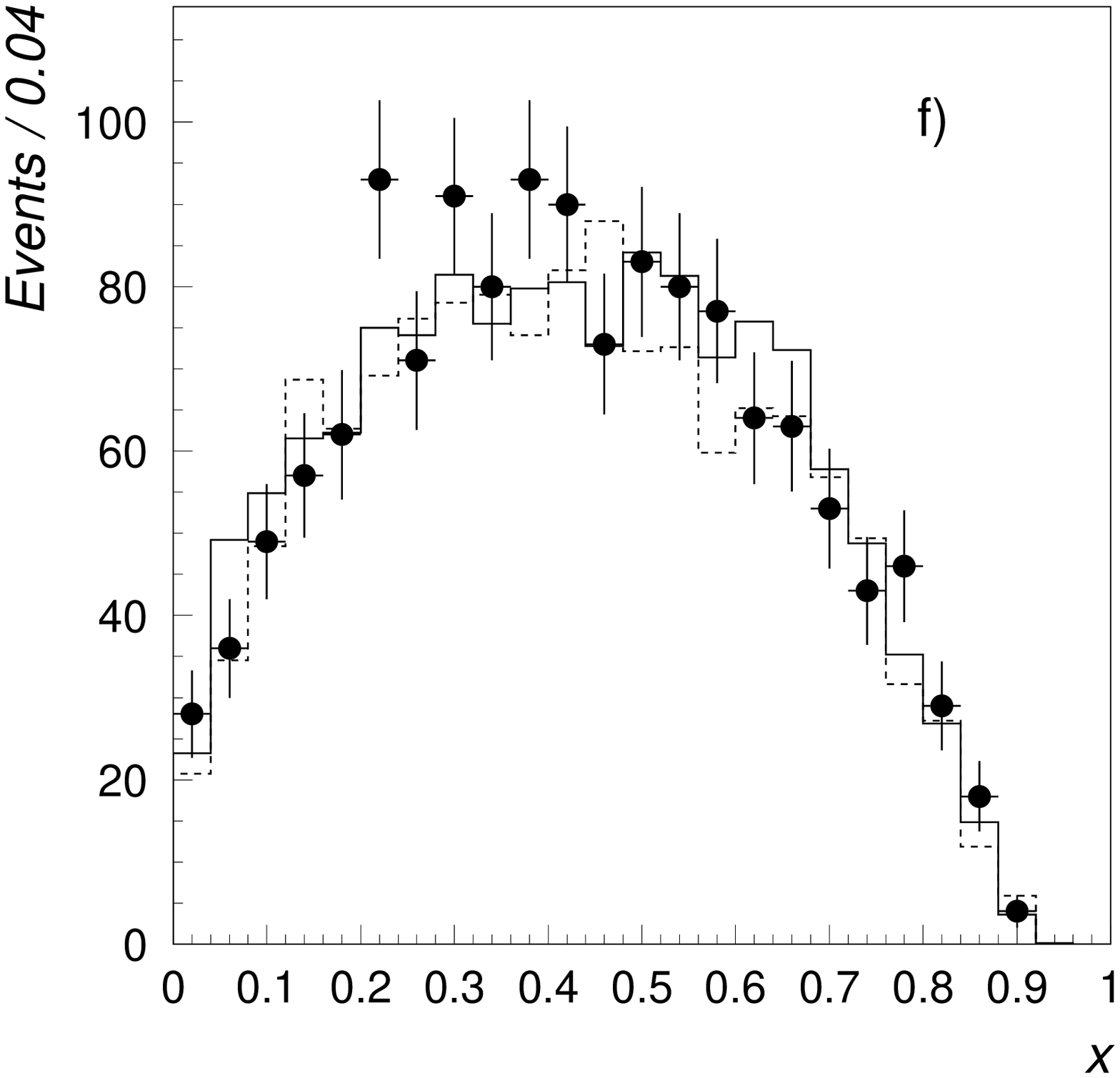,width=7cm }}
\end{center}
\caption{Distributions for the SAT single tagged events: a) 
$E_{tag}/E_{beam}$, b) $\theta_{tag}$ (180$^\circ -\theta_{tag}$ for
positrons), c) squared momentum transfer $Q^2$, d) invariant mass
of muon pair, e) sum of the transverse momenta of the muons, f) value of $x$.
The points correspond to the background subtracted data, the solid line
to the BDKRC simulation, and the dashed line to the DIAG36 simulation. }
\label{SAT}
\end{figure}

\begin{figure}
\begin{center}
{\Large DELPHI}
\mbox{\epsfig{file=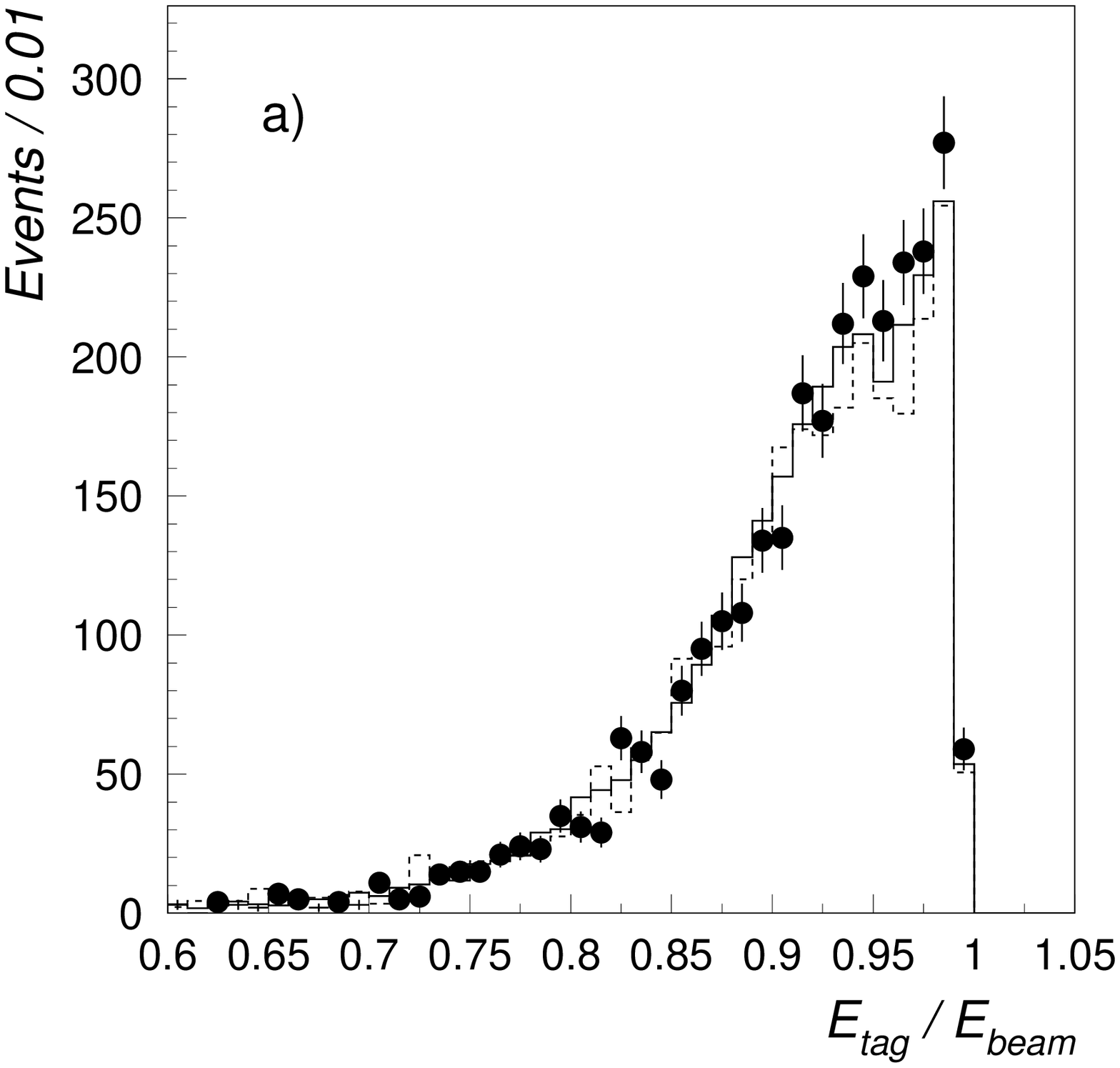,width=7cm }
      \epsfig{file=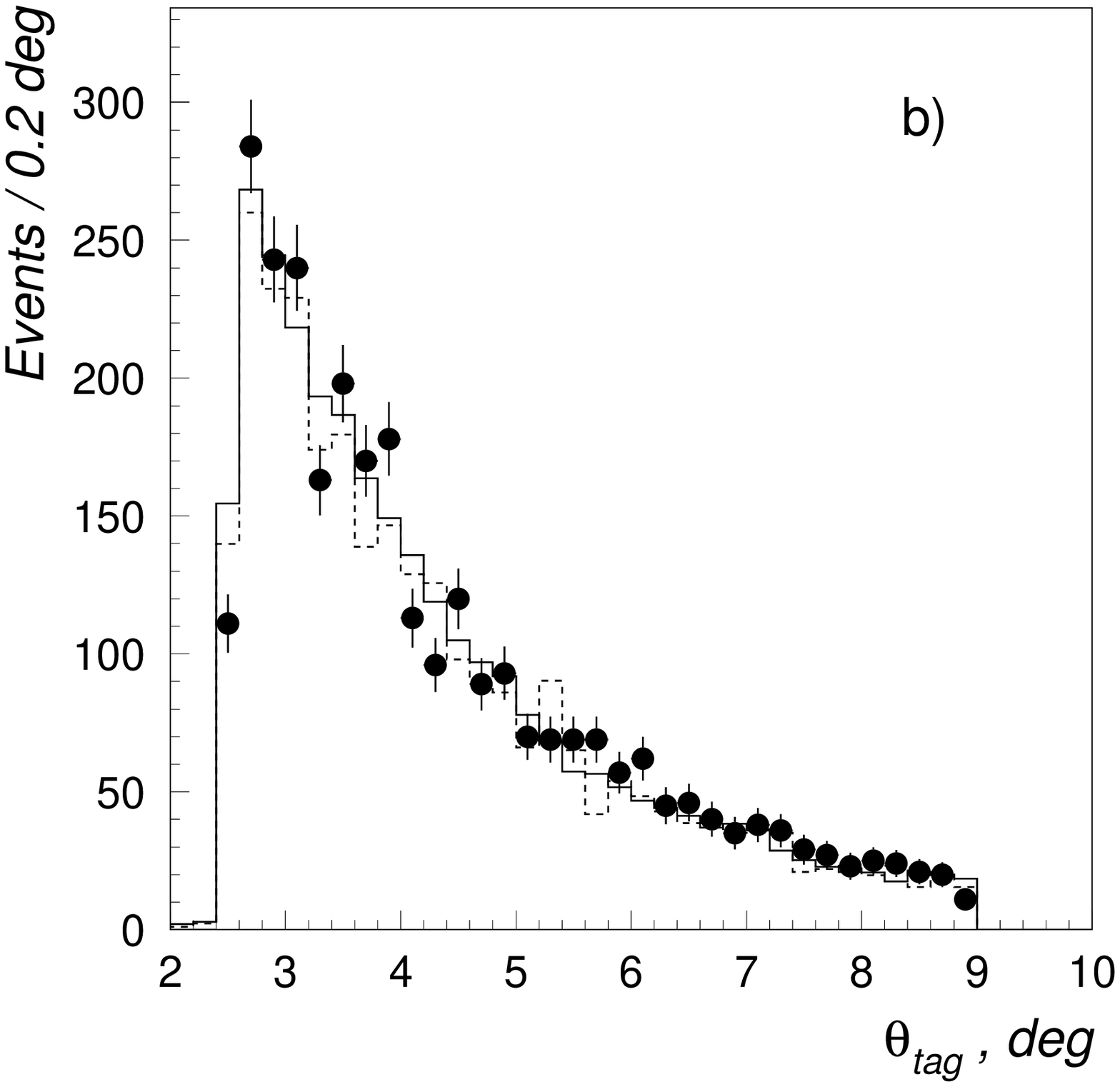,width=7cm }}
\end{center}
\vspace*{-1cm}
\begin{center}
\mbox{\epsfig{file=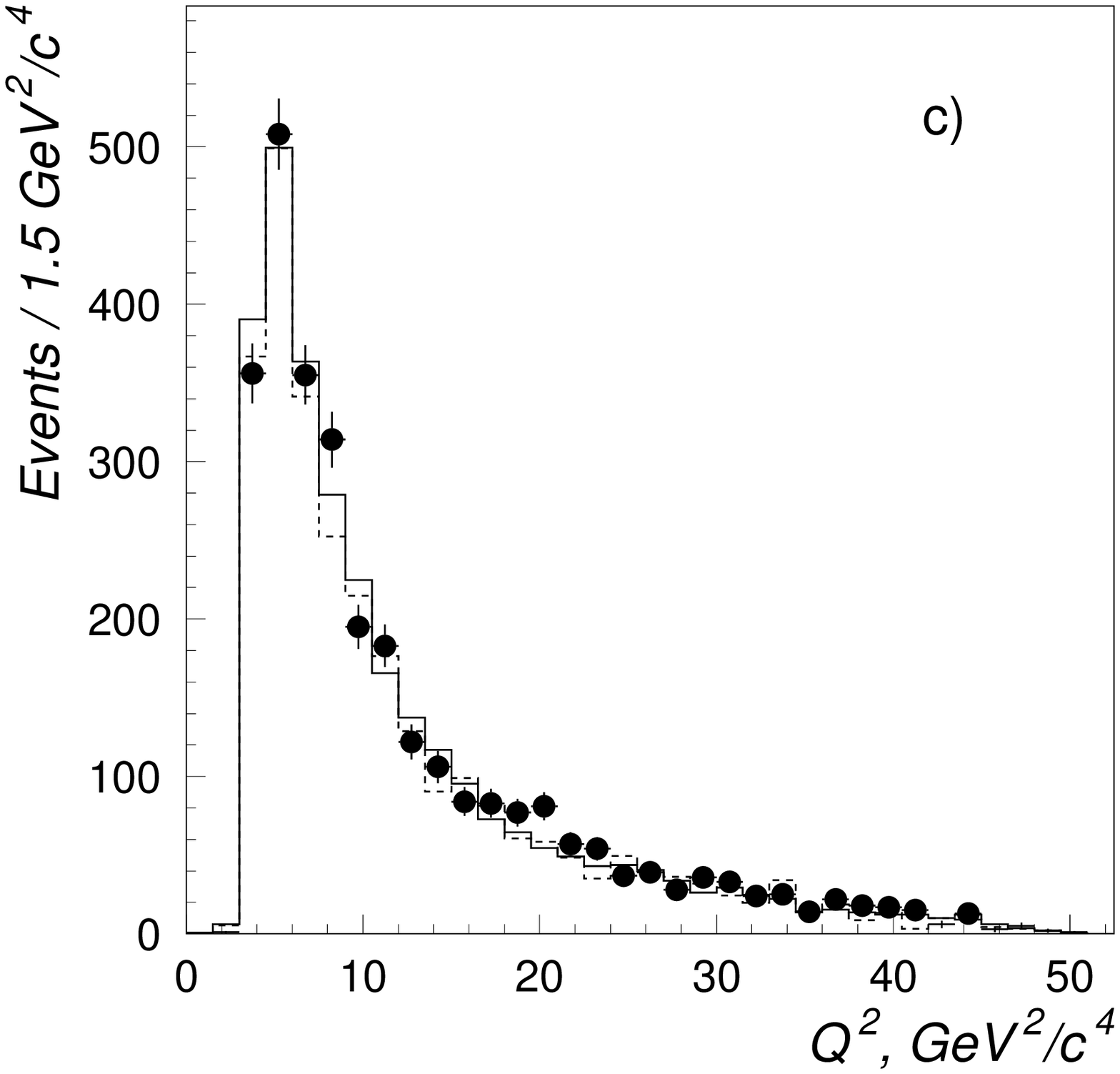,width=7cm }
      \epsfig{file=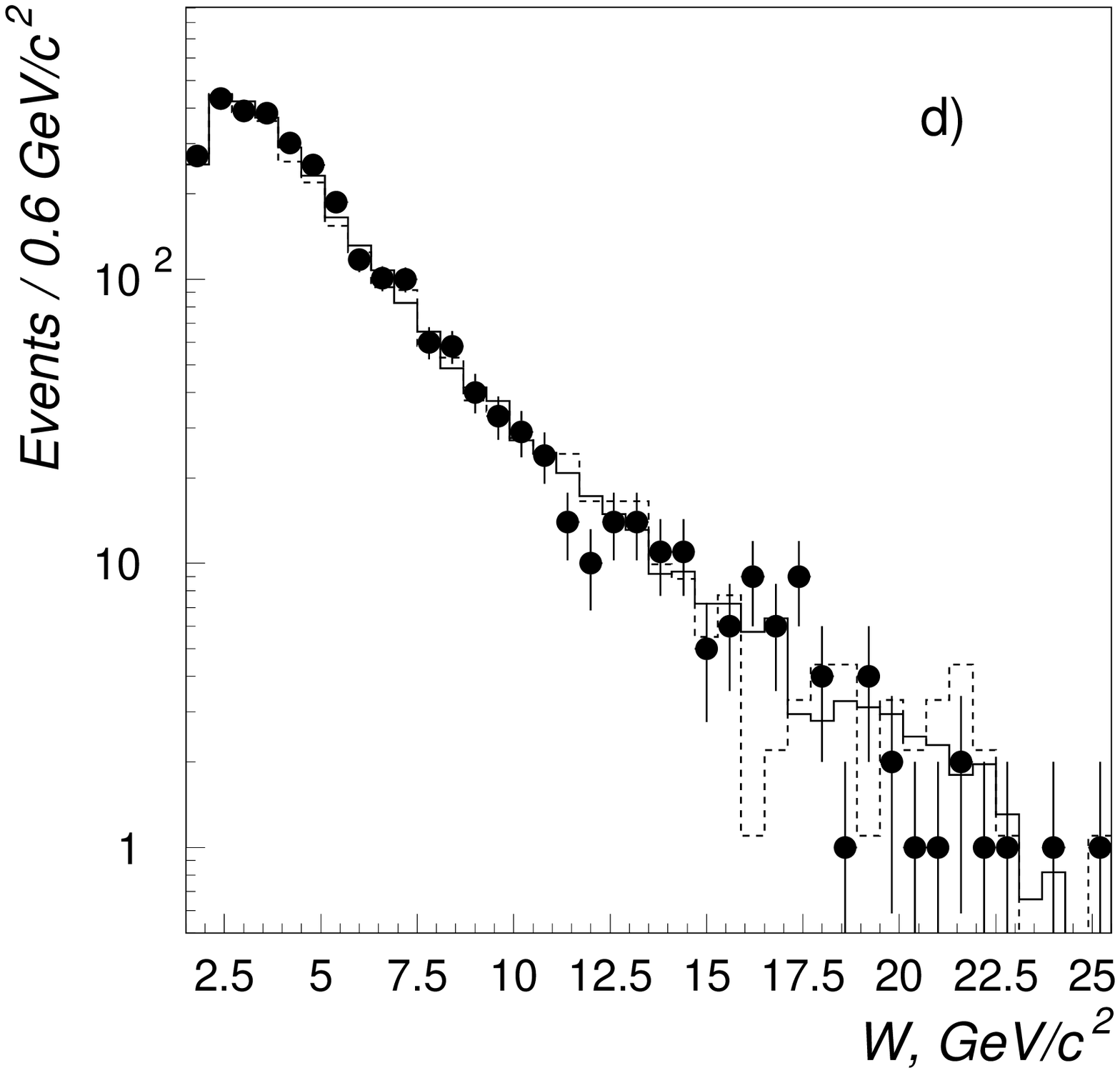,width=7cm }}
\end{center}
\vspace*{-1cm}
\begin{center}
\mbox{\epsfig{file=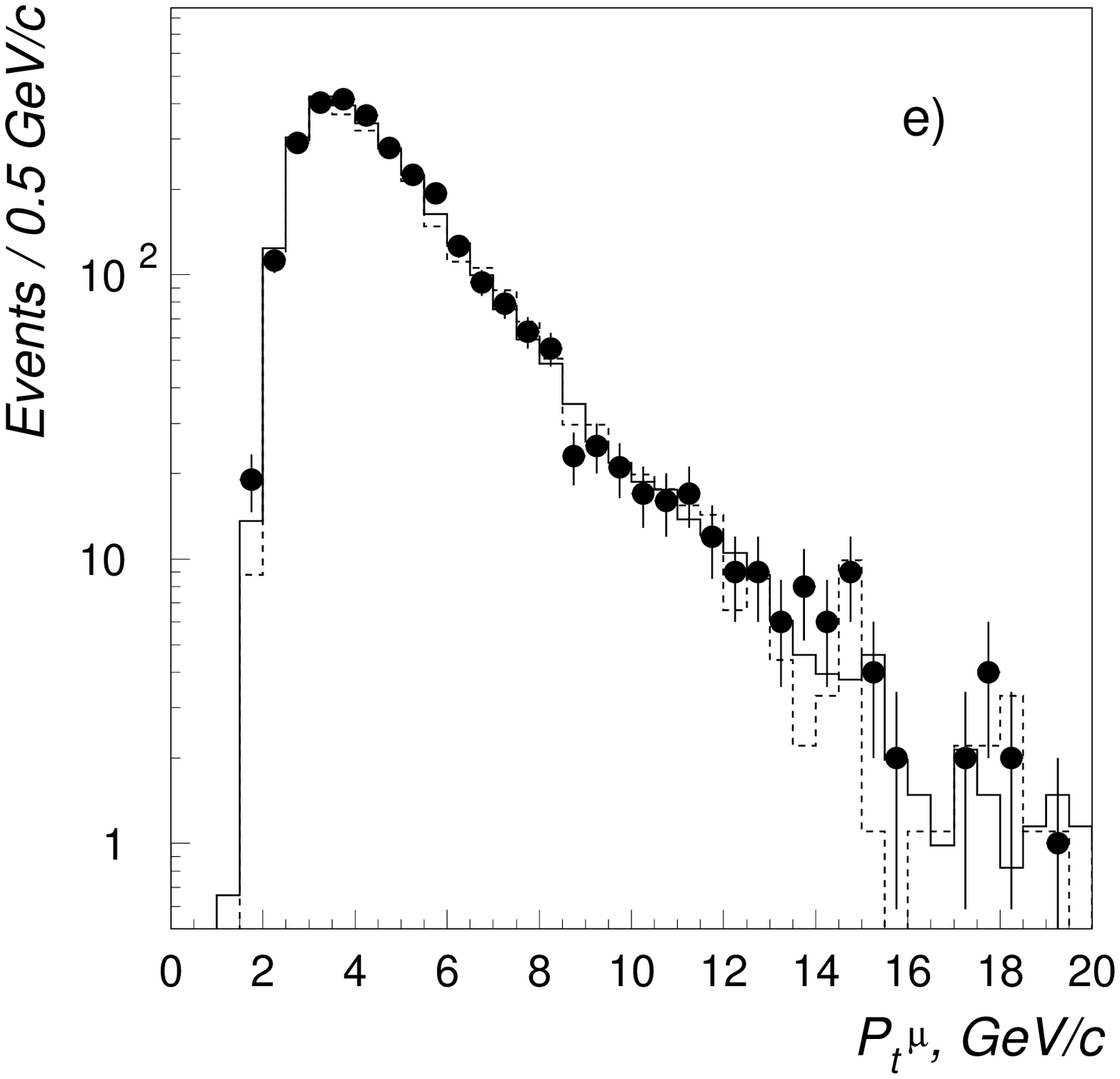,width=7cm }
      \epsfig{file=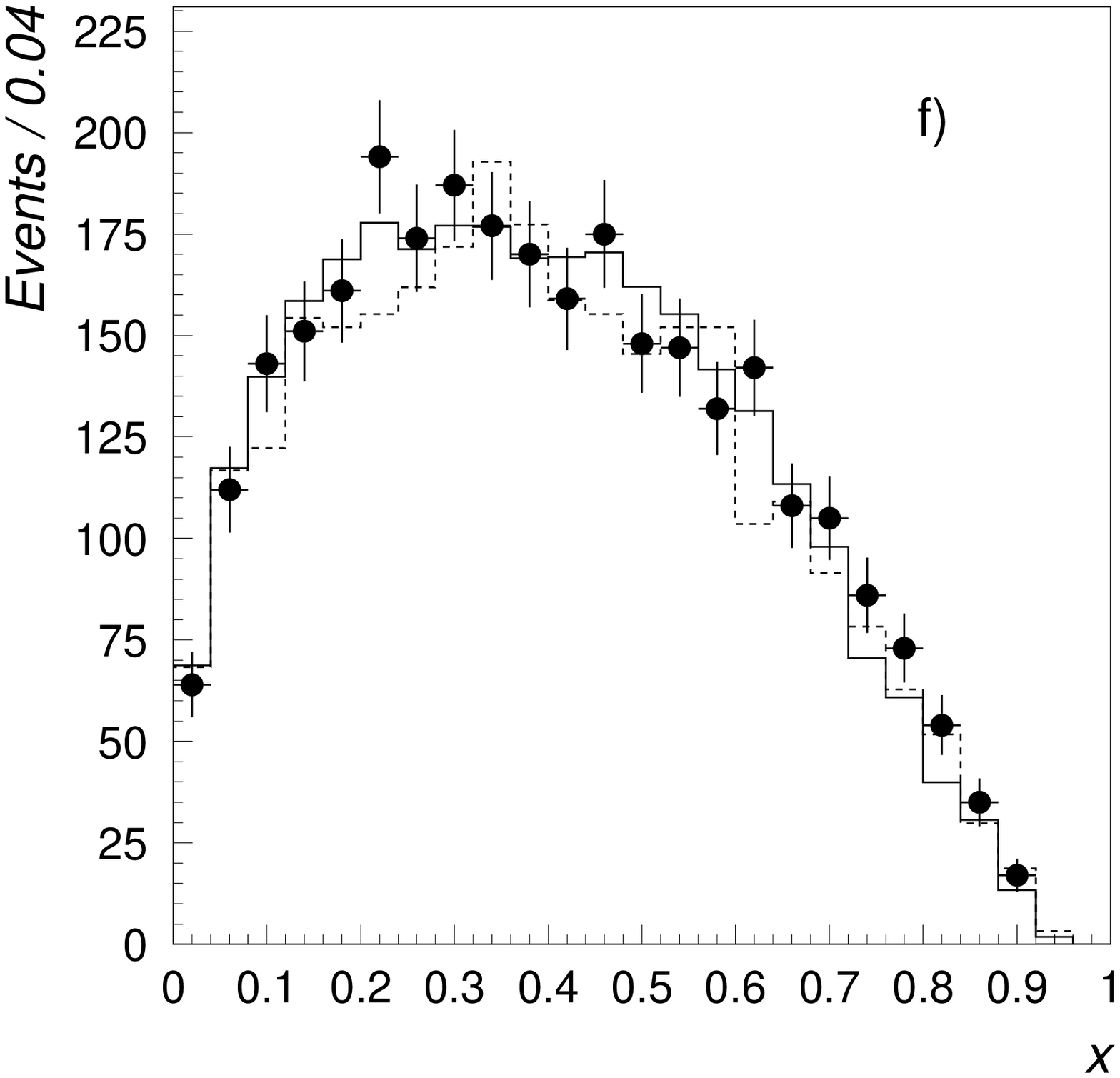,width=7cm }}
\end{center}
\caption{The same as Fig.~\ref{SAT} for the STIC single tagged events.}
\label{STIC}
\end{figure}
\clearpage

\begin{figure}
\begin{center}
{\Large DELPHI}
\mbox{\epsfig{file=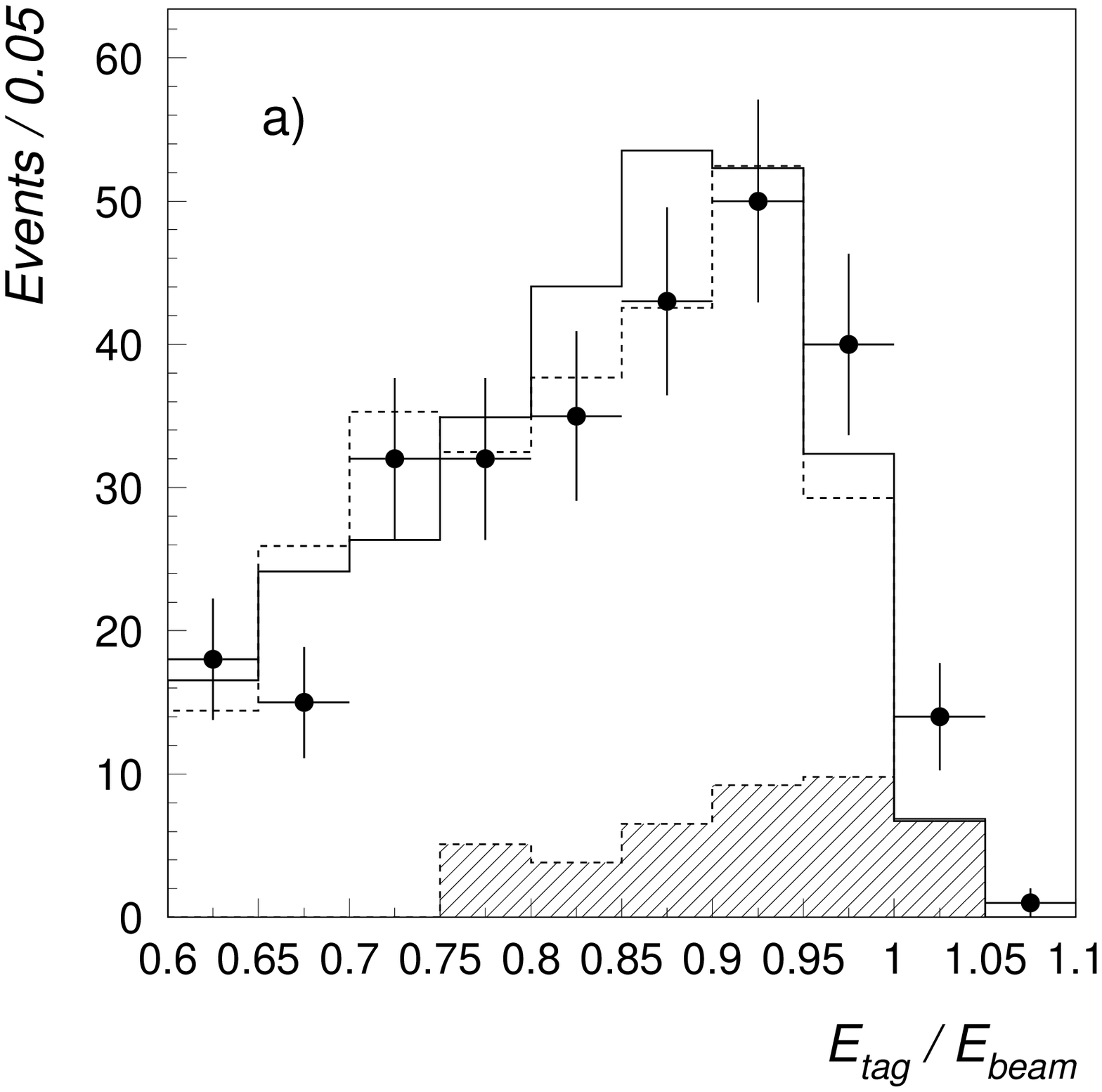,width=7cm }
      \epsfig{file=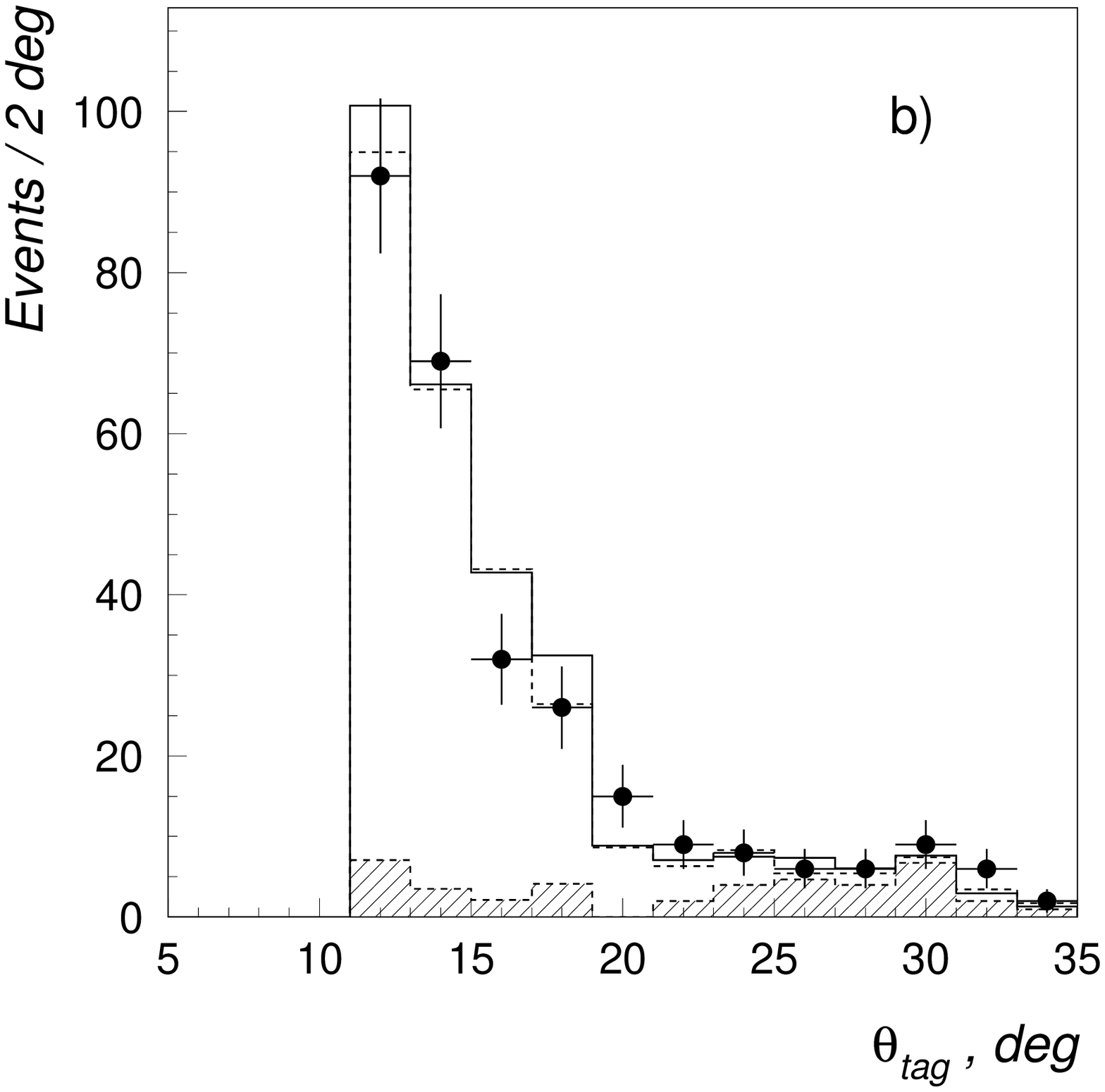,width=7cm }}
\end{center}
\vspace*{-1cm}
\begin{center}
\mbox{\epsfig{file=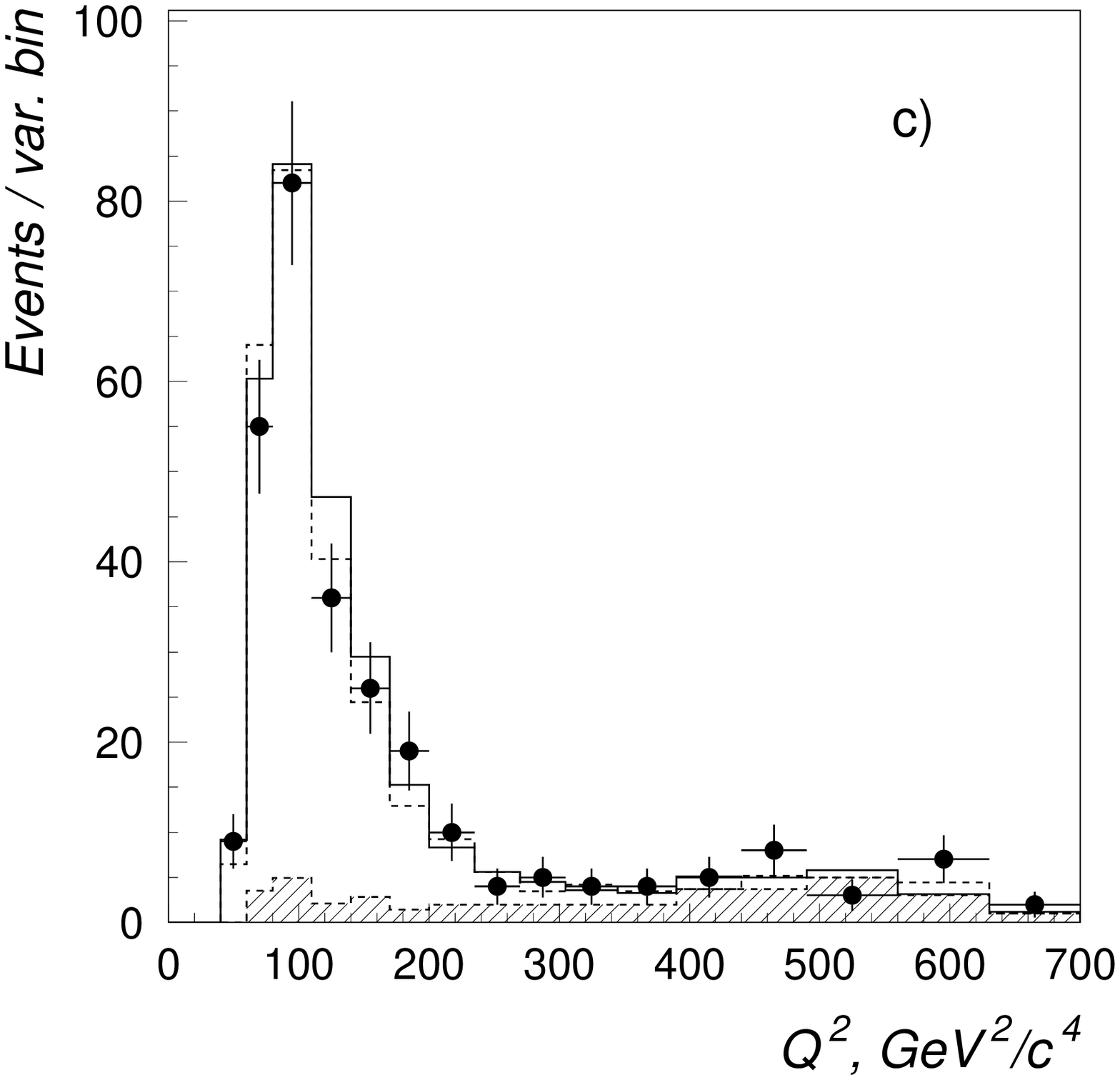,width=7cm }
      \epsfig{file=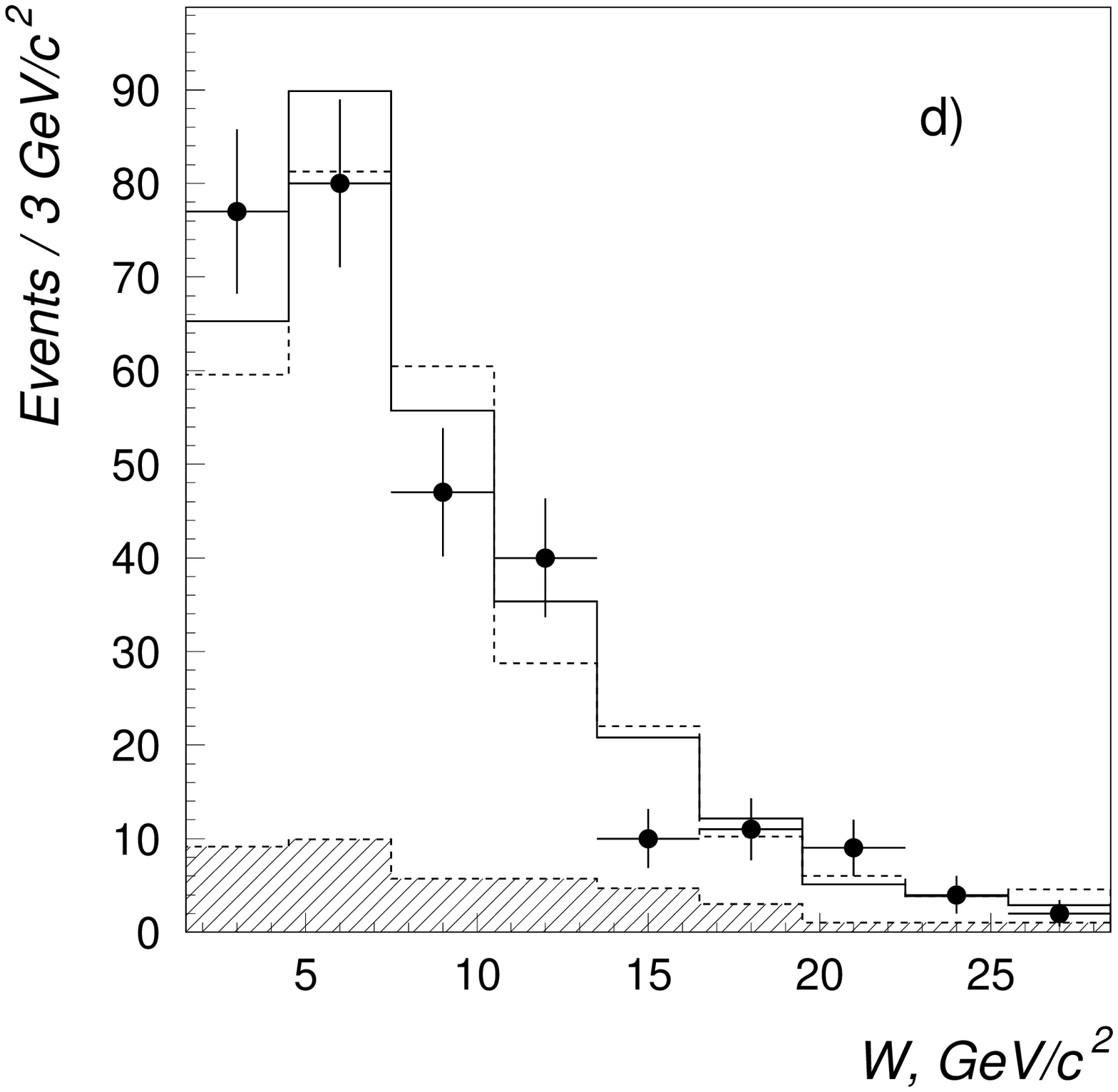,width=7cm }}
\end{center}
\vspace*{-1cm}
\begin{center}
\mbox{\epsfig{file=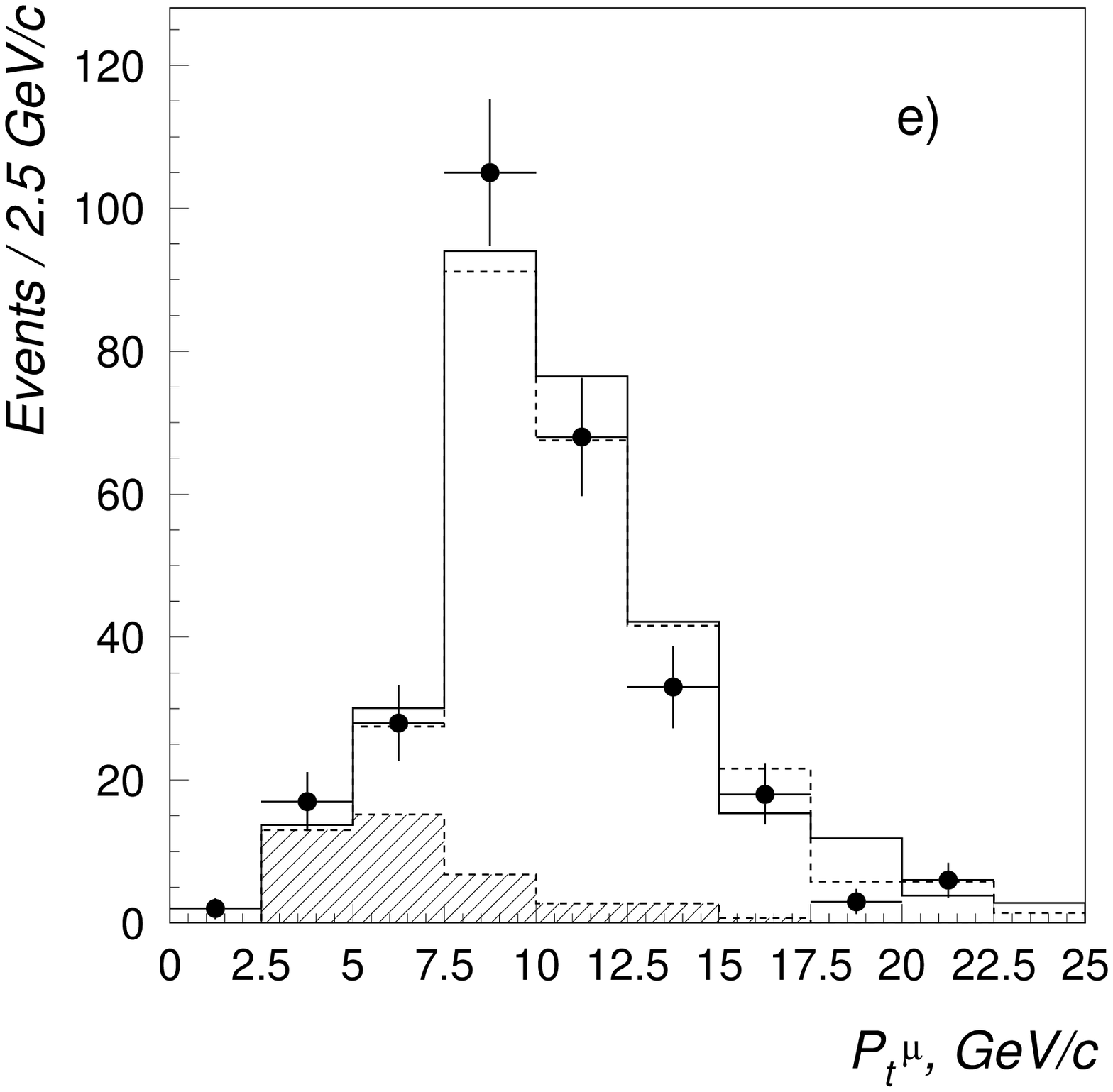,width=7cm }
      \epsfig{file=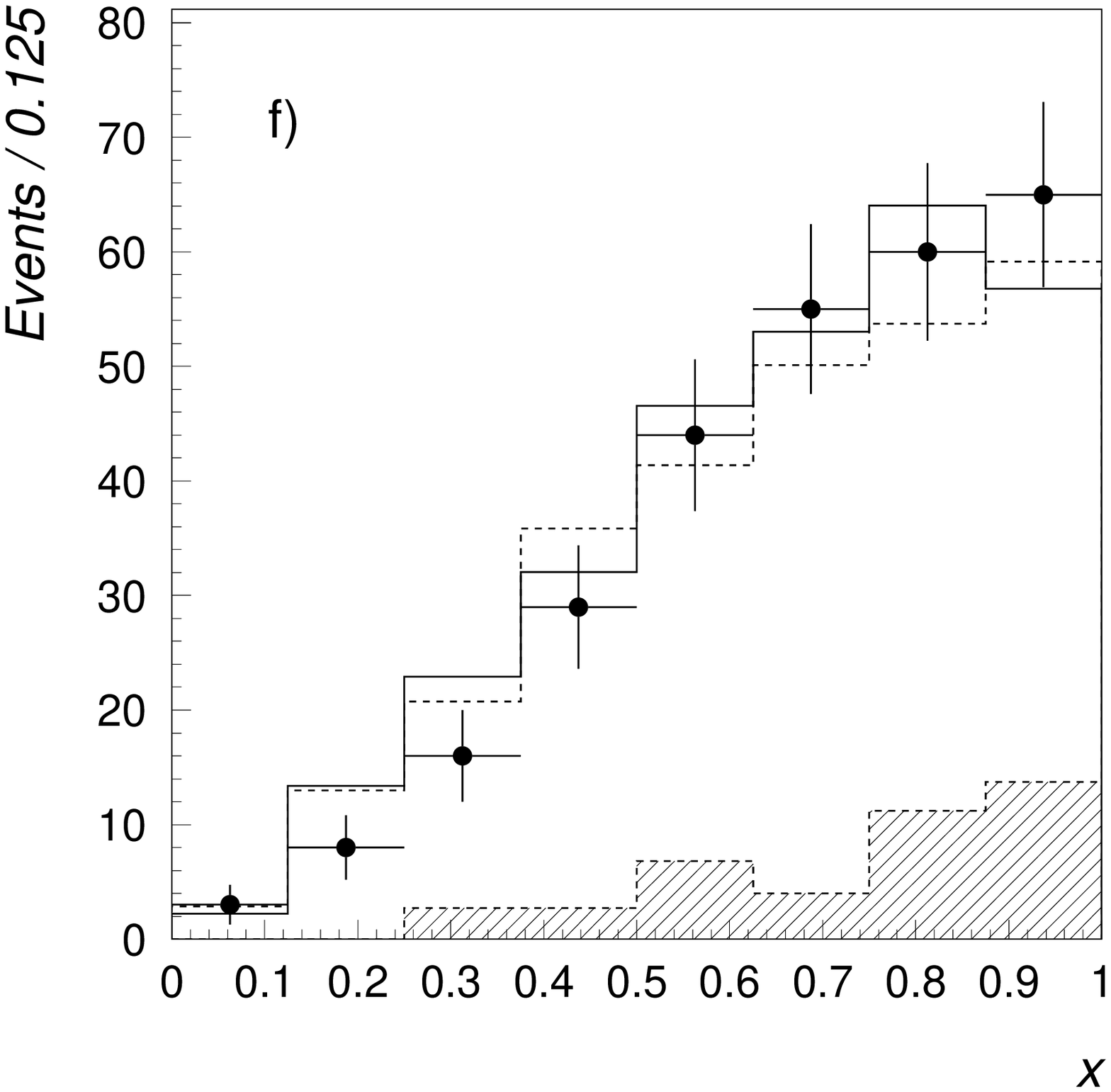,width=7cm }}
\end{center}
\caption{The same as Figs.~\ref{SAT} and \ref{STIC} for the FEMC single tagged
events except that the background, relatively much larger here than in 
Figs.~\ref{SAT} and \ref{STIC} and shown here by the hatched histograms, 
has in this case been added to the simulated distributions rather than
subtracted from the data.}
\label{FEMC}
\end{figure}
\clearpage

\begin{figure}[t]
\begin{center}
{\Large DELPHI}\\
\mbox{\epsfig{file=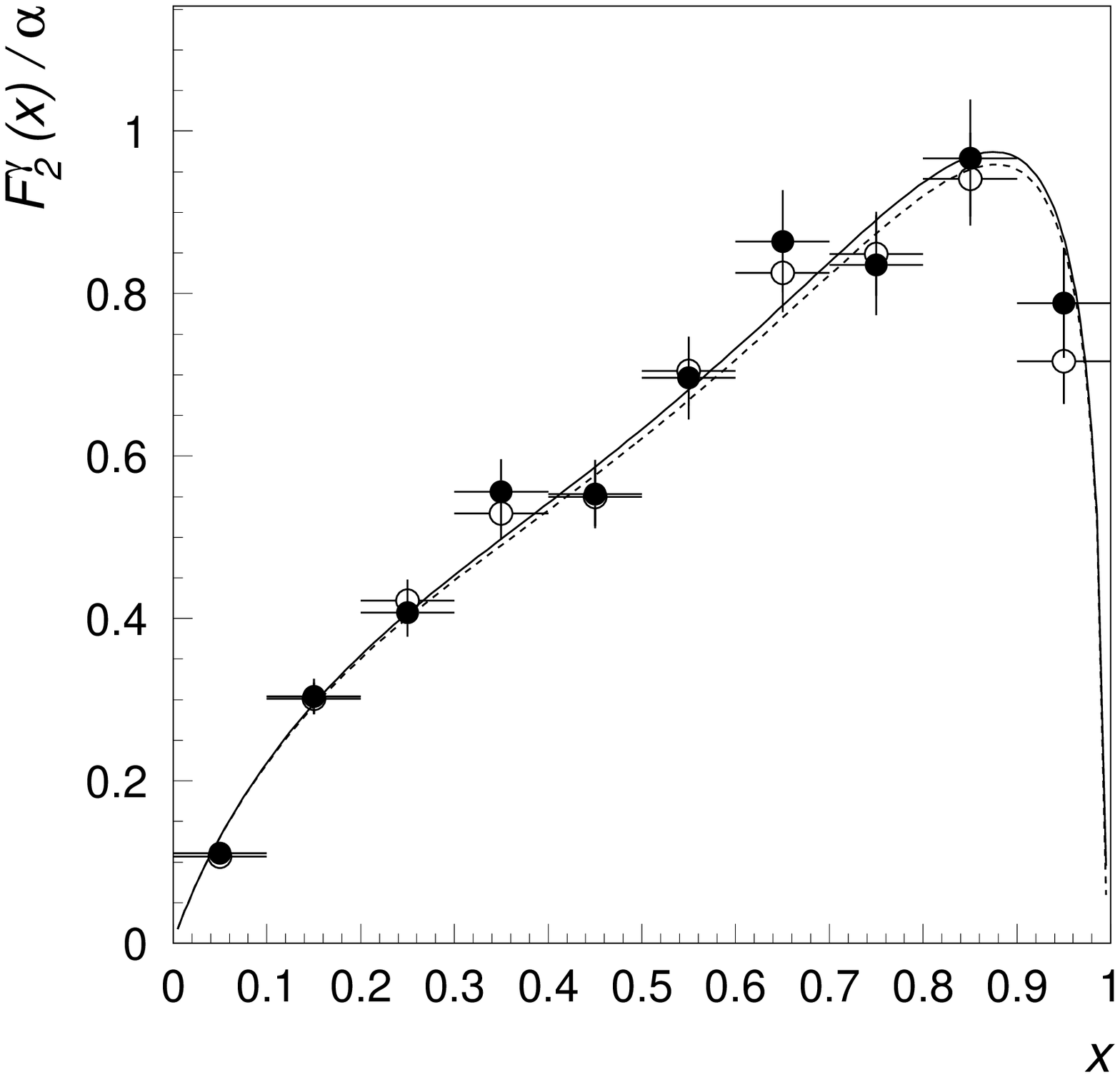,width=8cm }}
\end{center}
\vspace*{-7mm}
\caption[]{$F_2^\gamma(x)$ for $<Q^2>=12.5$ GeV$^2/c^4$ extracted from 
a simulated STIC tagged event sample (points) and the fit to the QED expression
(lines): full circles and solid lines are for the extraction using the
simulation with known $F_2$, open circles and dashed lines for the
extraction using the photon flux approach. }
\label{F2gen}
\end{figure}

\begin{figure}[hb]
\begin{center}
{\Large DELPHI}\\
\mbox{\epsfig{file=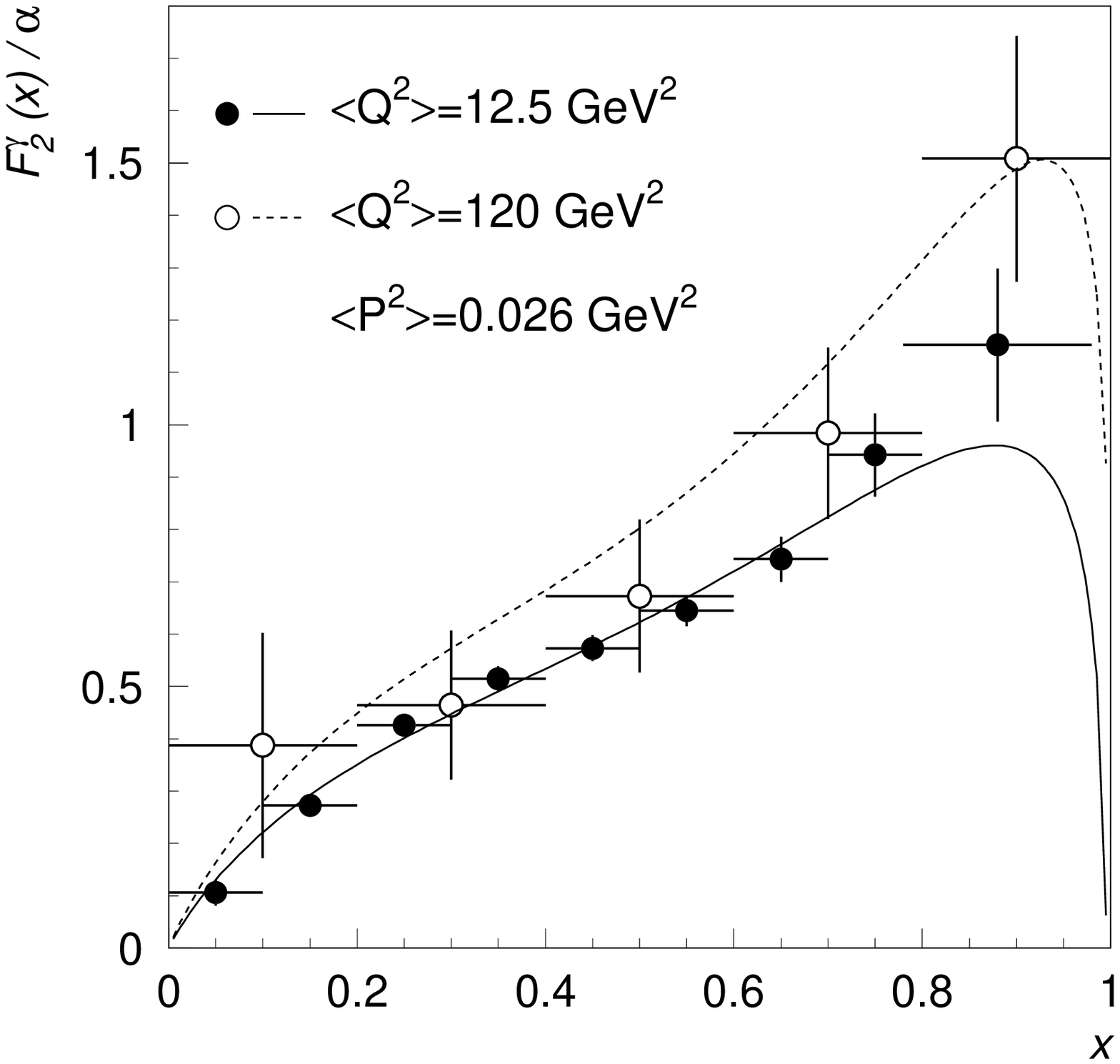,width=8cm }}
\end{center}
\vspace*{-7mm}
\caption[]{$F_2^\gamma(x)$ extracted from the combined SAT and STIC data 
($<Q^2>=12.5$ GeV$^2/c^4$, full circles), and from the FEMC data 
($<Q^2>=120$ GeV$^2/c^4$, open circles). 
Statistical and systematic errors are added in quadrature. 
The solid and dashed lines show the QED predictions
with $P^2$=0.026~GeV$^2/c^4$ for the low $Q^2$ and high $Q^2$ samples,
respectively.}
\label{F2exp}
\end{figure}
\clearpage

\begin{figure}
\begin{center}
{\Large DELPHI}
\mbox{\epsfig{file=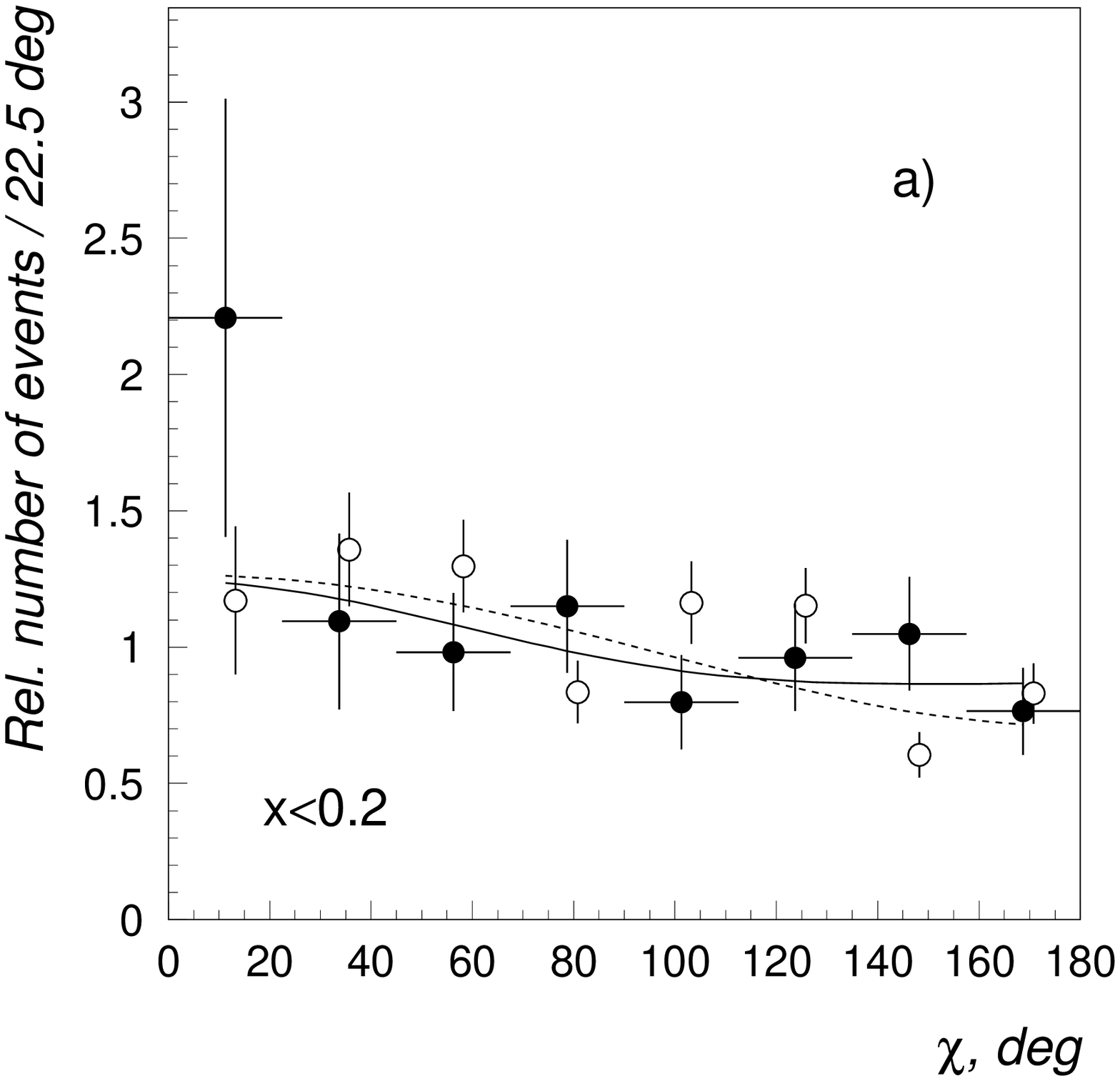,width=7cm }
      \epsfig{file=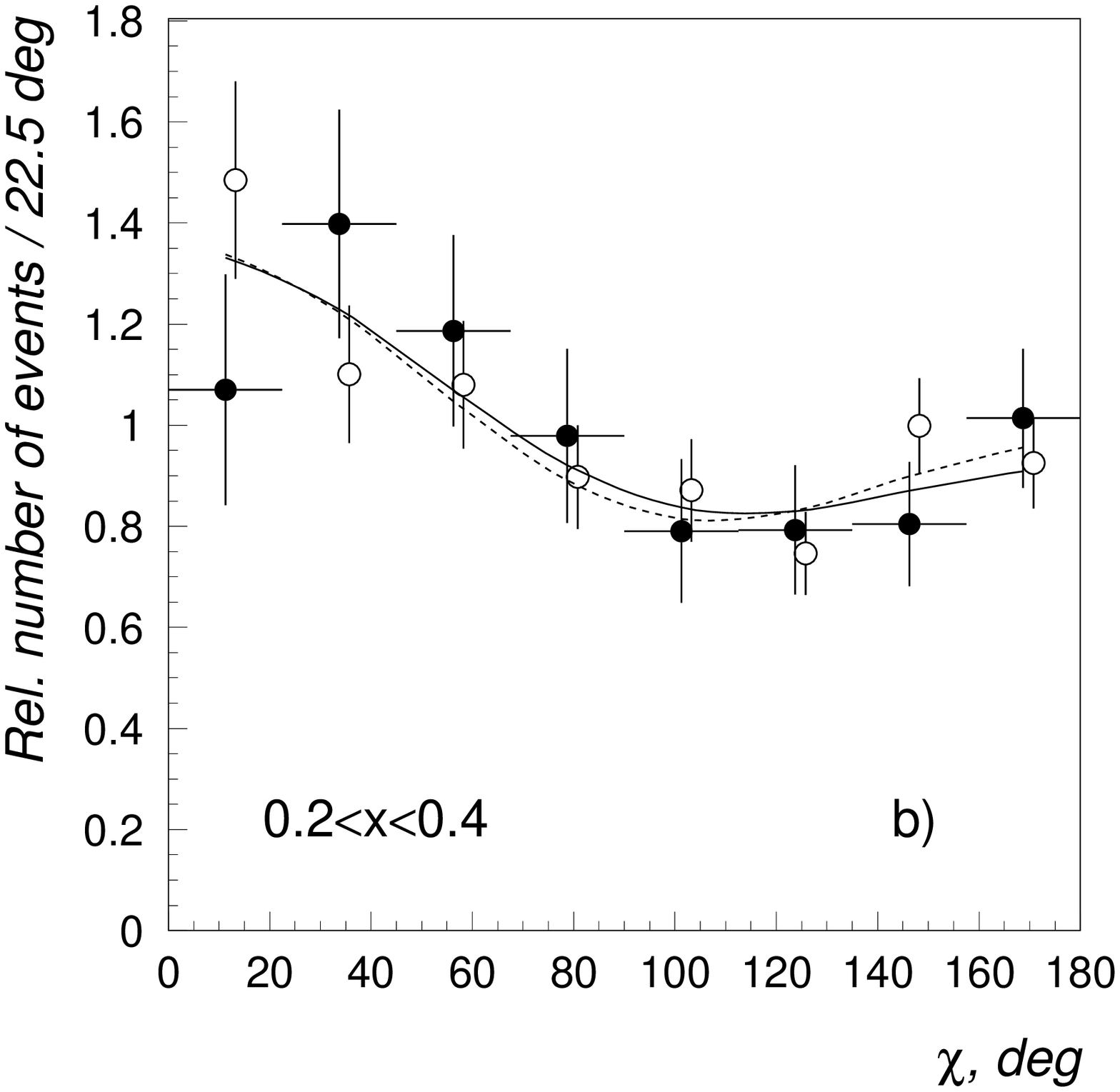,width=7cm }}
\end{center}
\vspace*{-1cm}
\begin{center}
\mbox{\epsfig{file=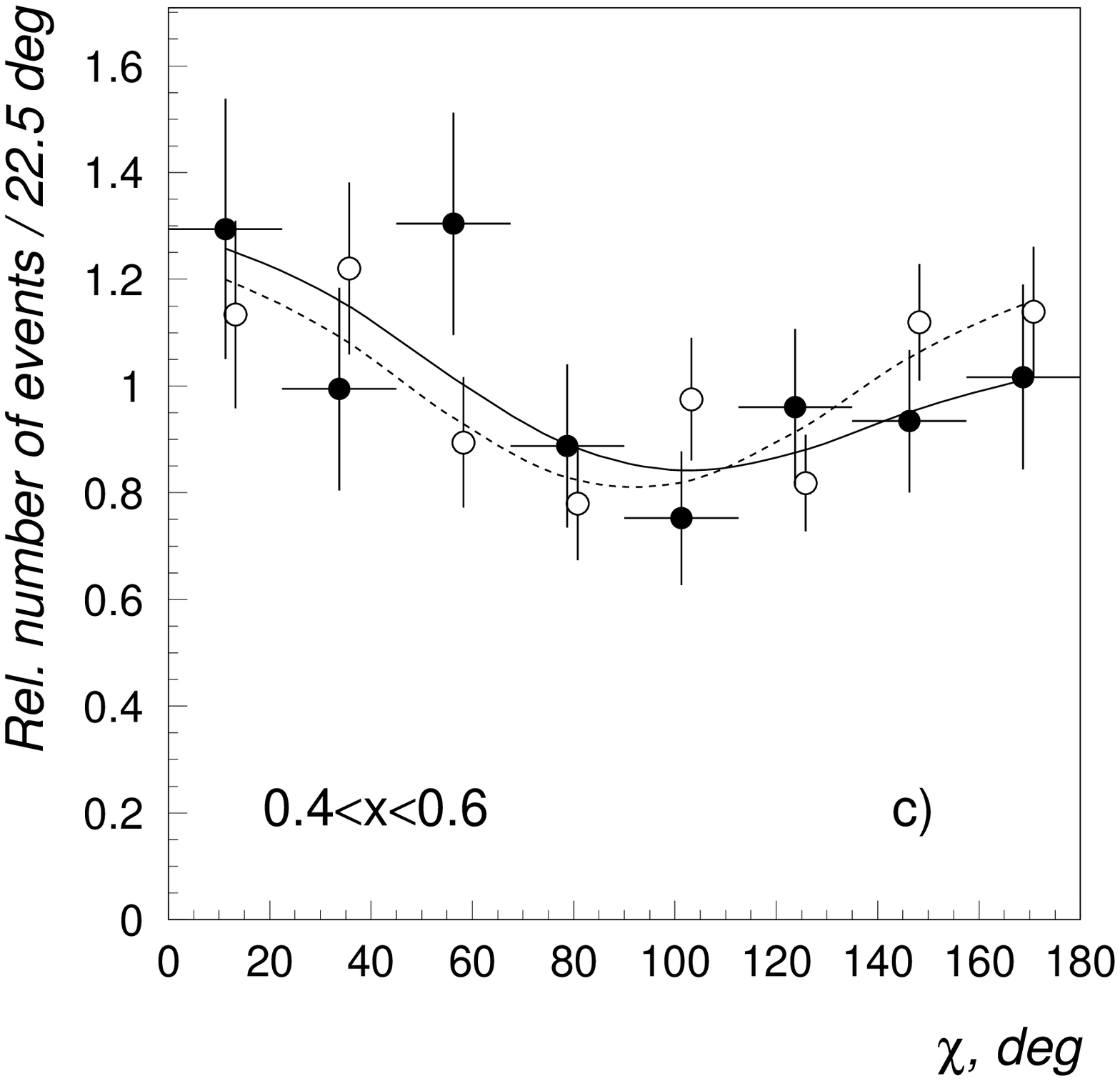,width=7cm }
      \epsfig{file=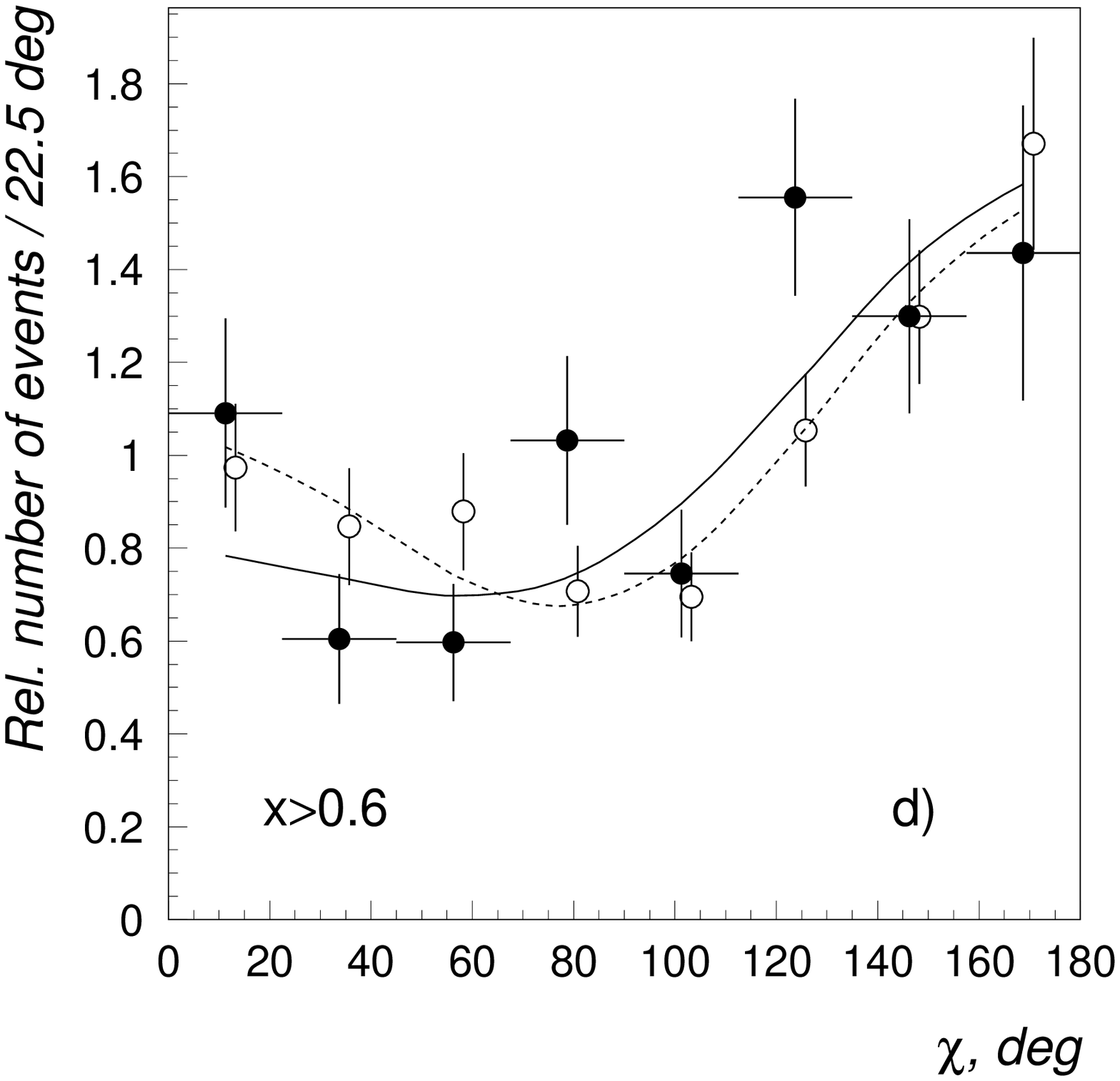,width=7cm }}
\end{center}
\vspace*{-1cm}
\begin{center}
\mbox{\epsfig{file=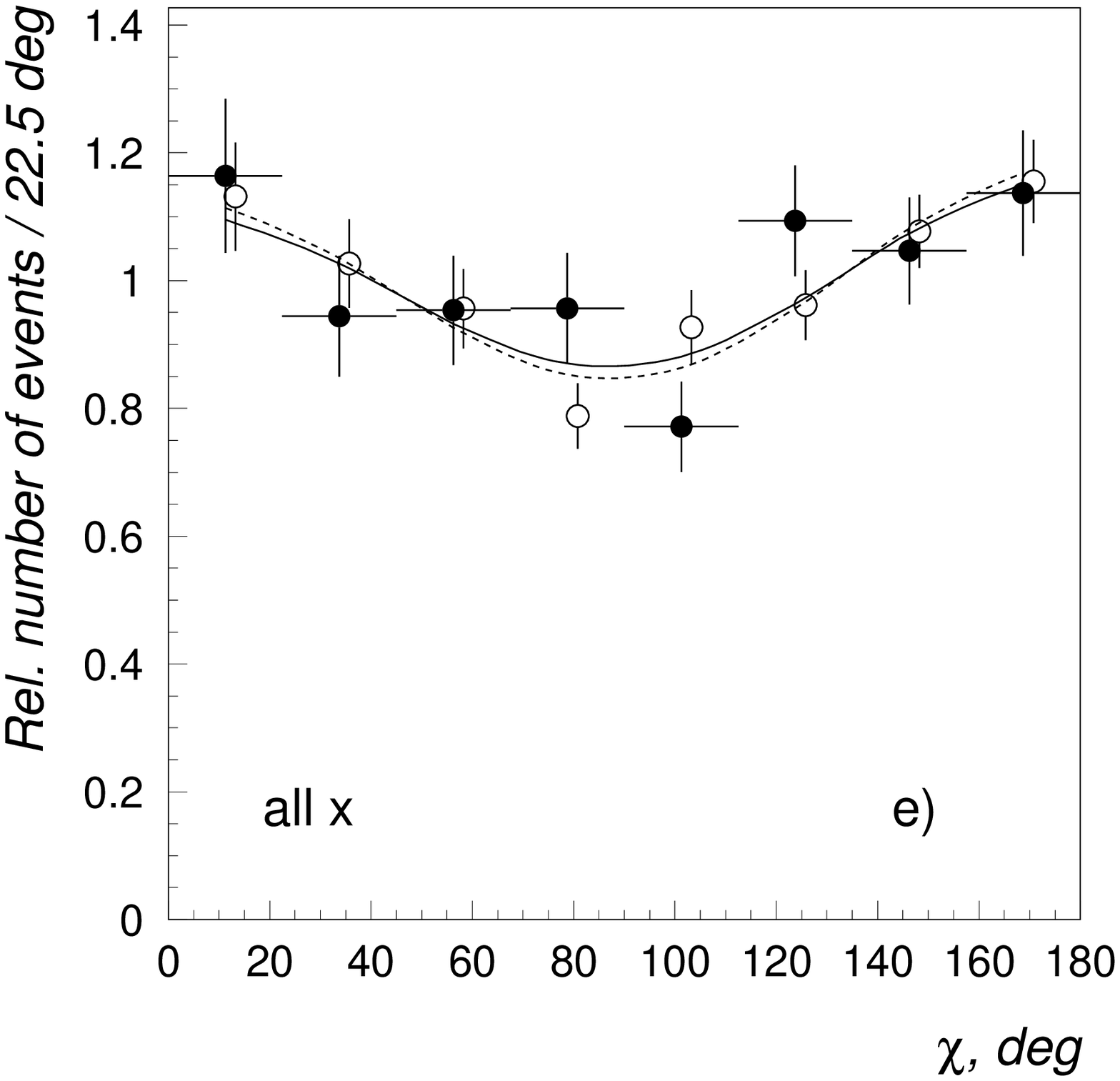,width=7cm }}
\end{center}
\caption{Azimuthal angle distributions corrected for detector inefficiency:
a) $x<$0.2, b) 0.2$<x<$0.4, c) 0.4$<x<$0.6, d) $x>$0.6, e)
all $x$. The lines correspond to the results of the fit. Full circles 
and solid line are for the SAT single tagged events, open circles 
and dashed line for the STIC events. }
\label{cor}
\end{figure}

\begin{figure}
\begin{center}
{\Large DELPHI}
\mbox{\epsfig{file=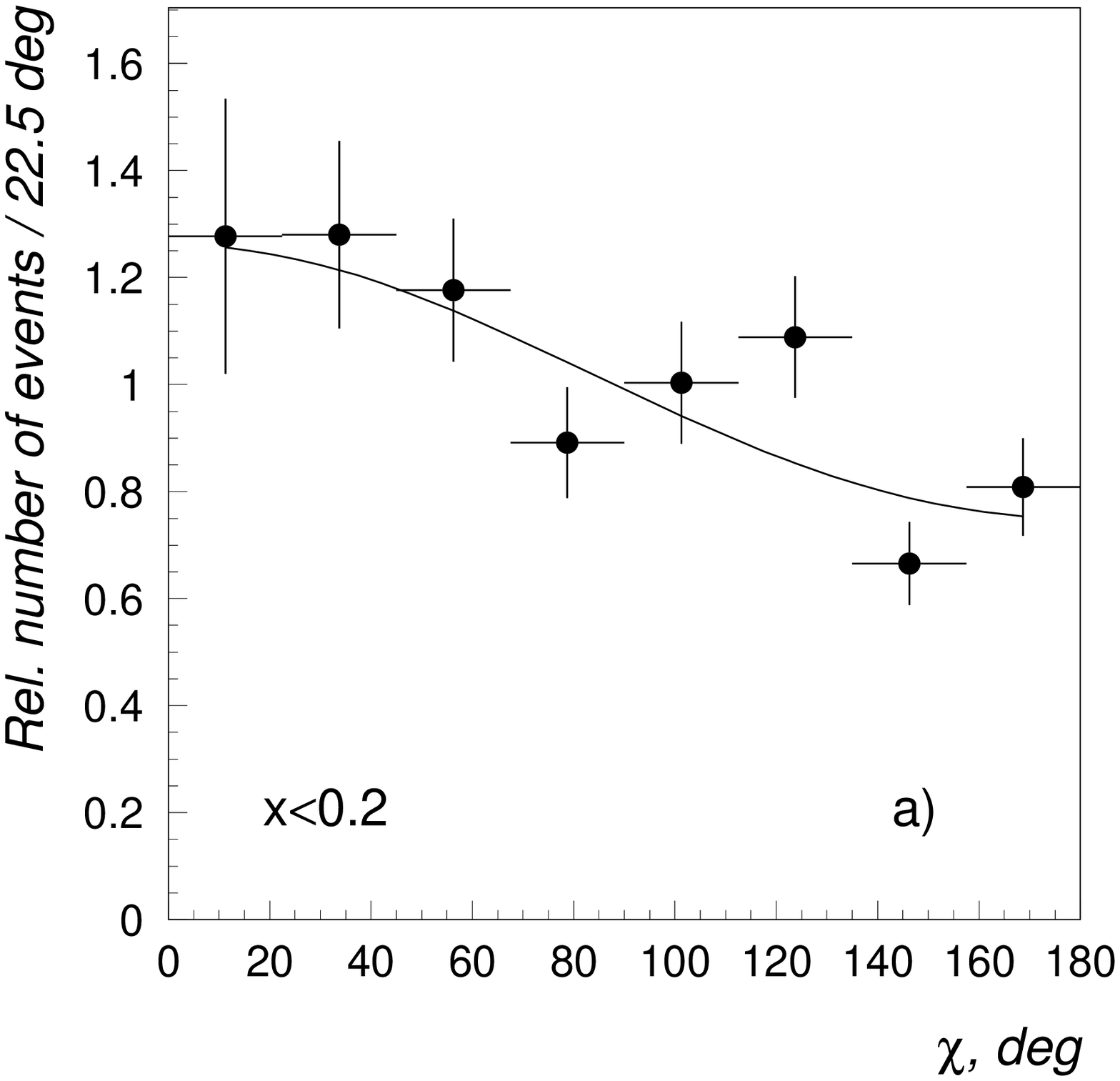,width=7cm }
      \epsfig{file=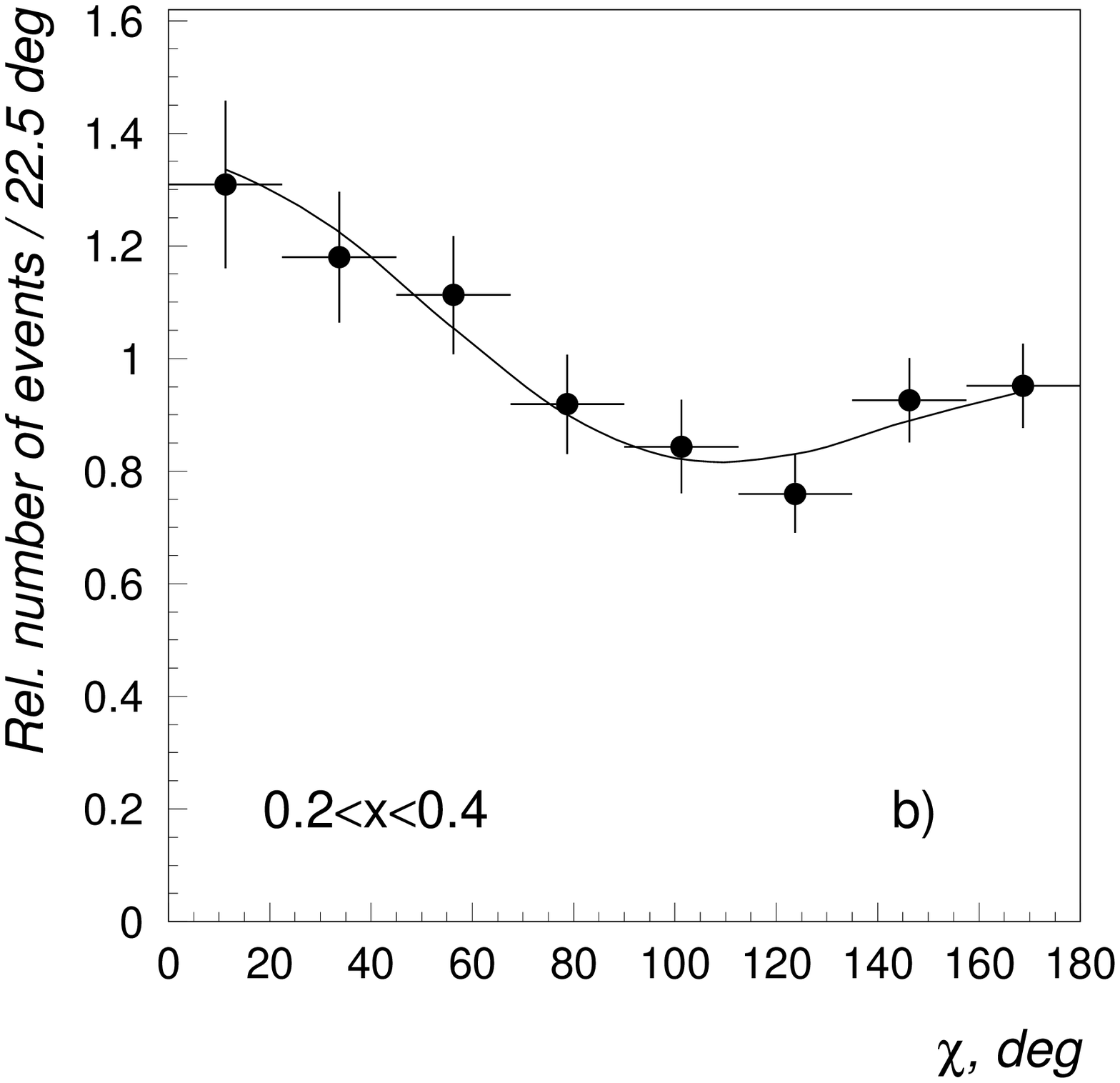,width=7cm }}
\end{center}
\vspace*{-1cm}
\begin{center}
\mbox{\epsfig{file=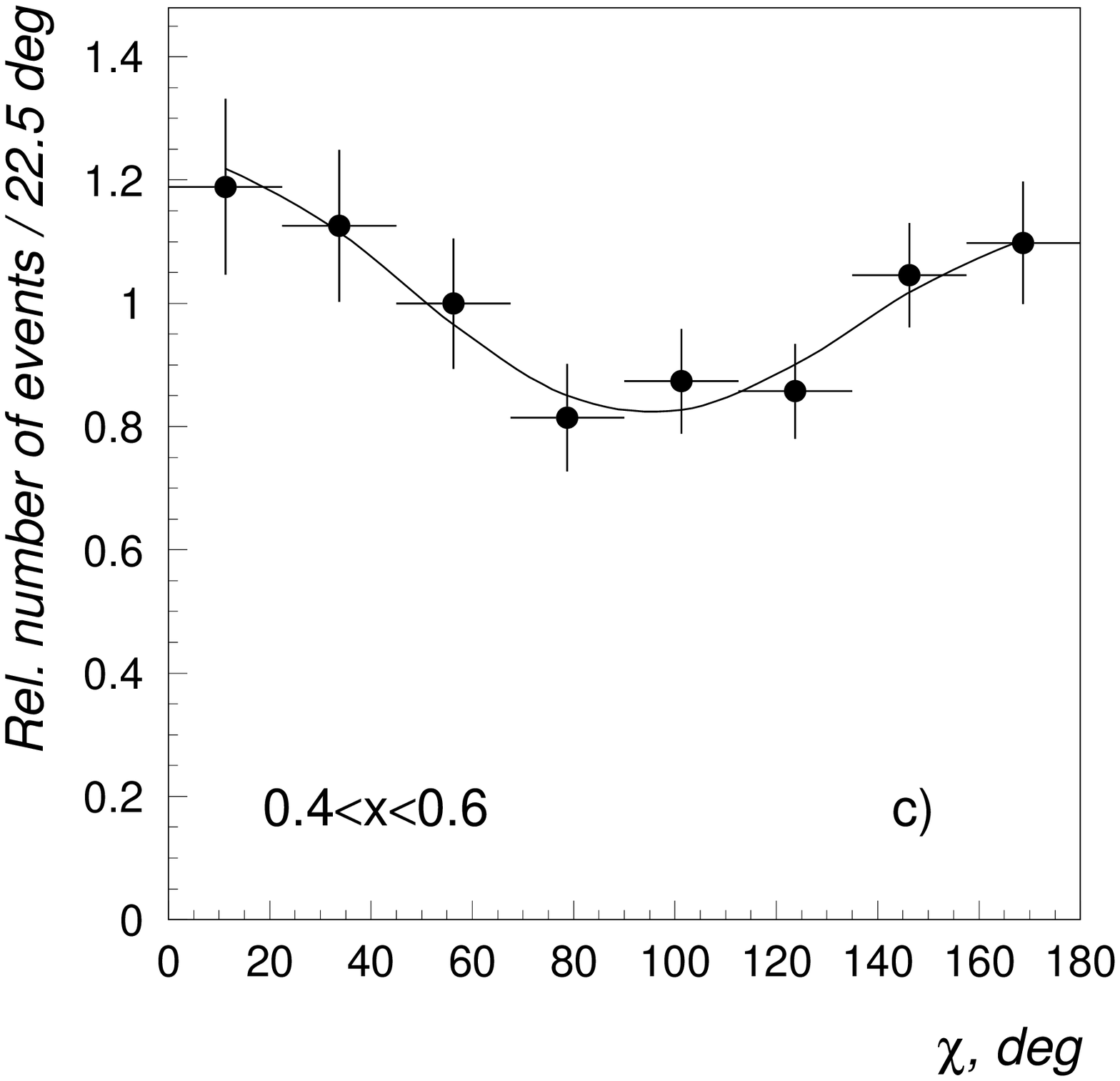,width=7cm }
      \epsfig{file=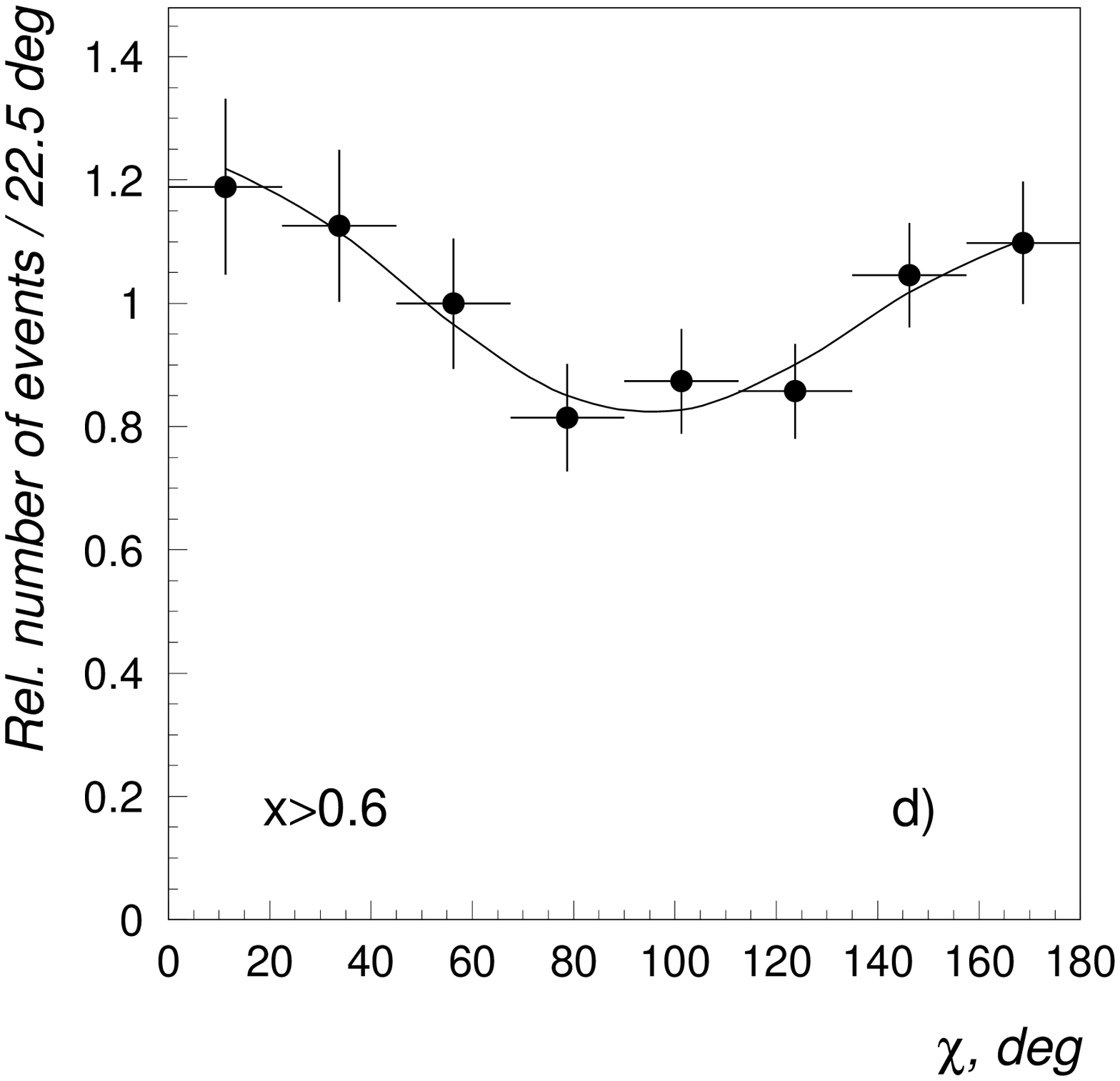,width=7cm }}
\end{center}
\vspace*{-1cm}
\begin{center}
\mbox{\epsfig{file=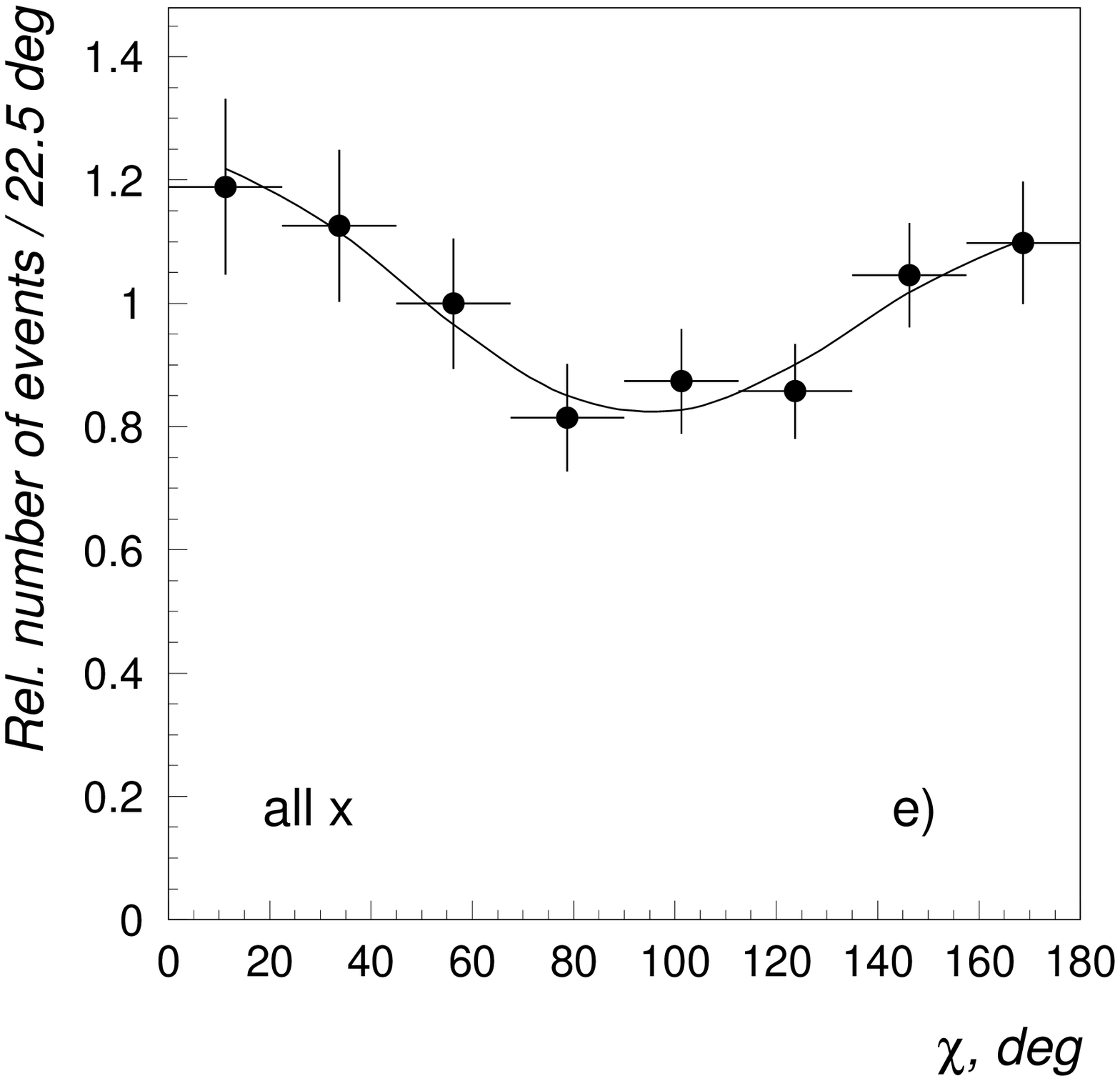,width=7cm }}
\end{center}
\caption{The same as Fig.~\ref{cor} for the combined SAT and STIC samples. }
\label{cor_comb}
\end{figure}

\begin{figure}
\begin{center}
{\Large DELPHI}
\mbox{\epsfig{file=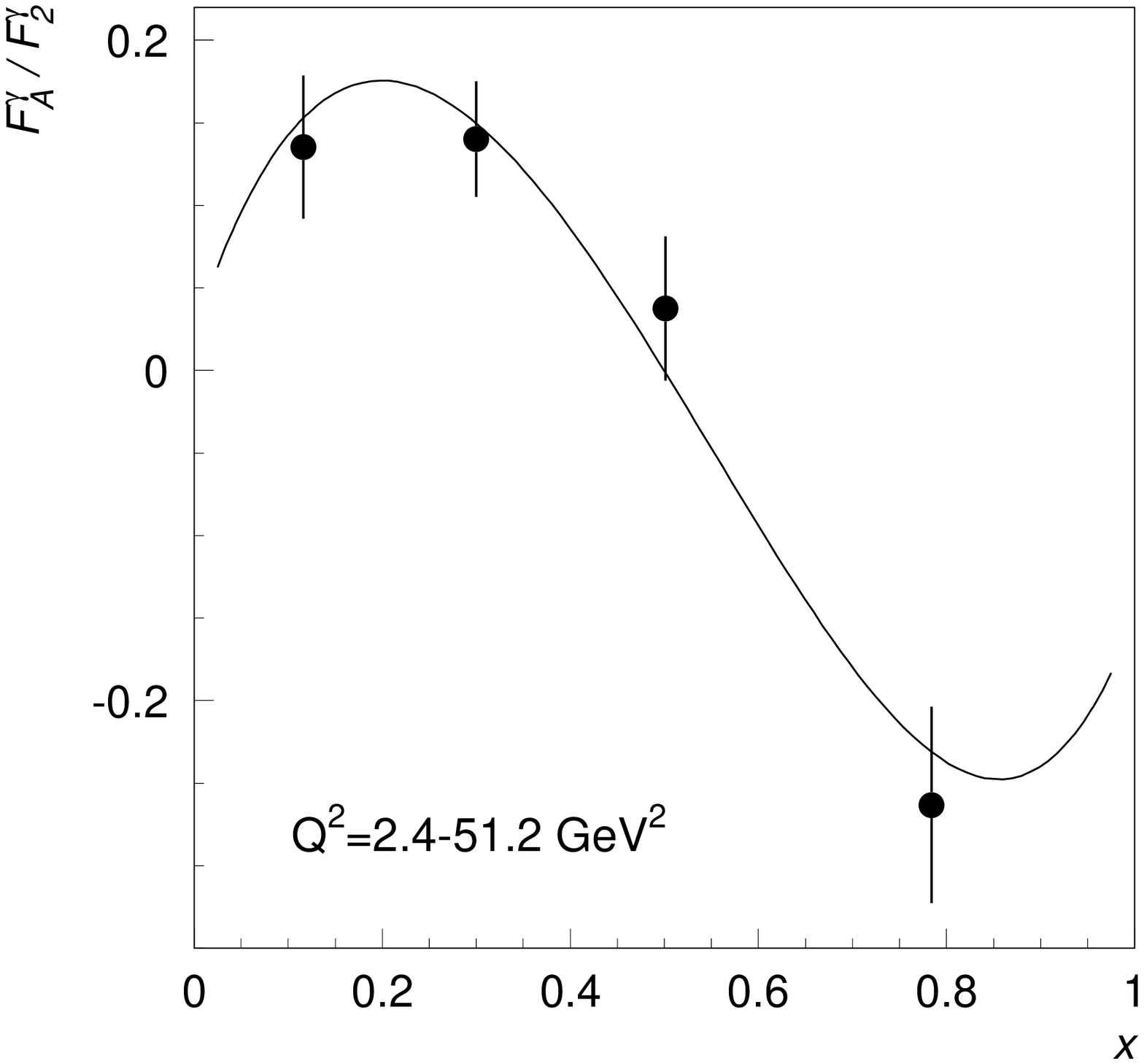,width=8cm }
      \epsfig{file=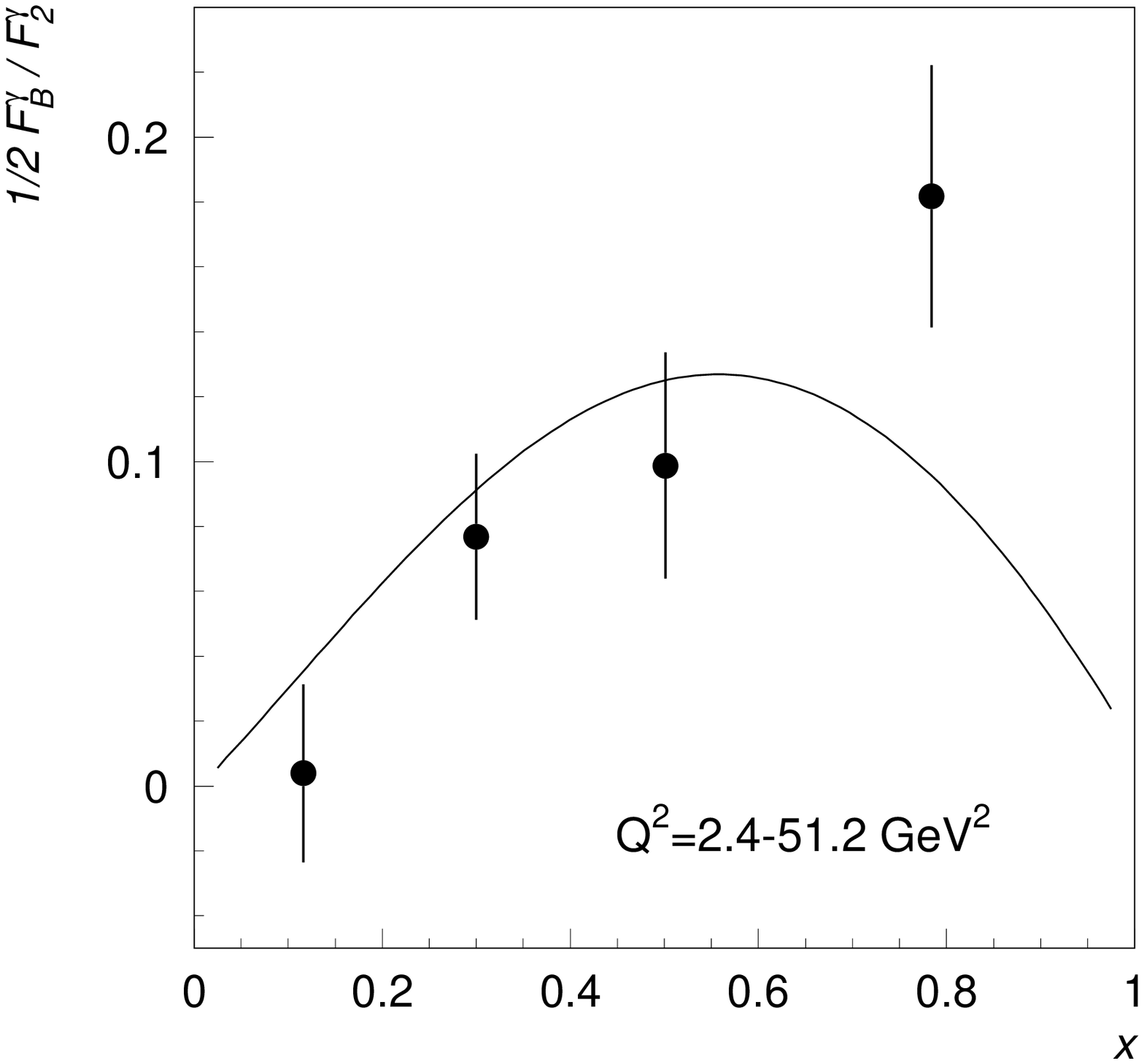,width=8cm }}
\end{center}
\caption[]{Ratios of leptonic structure functions $F_A^\gamma / F_2^\gamma$
(left) and ${1\over2}F_B^\gamma / F_2^\gamma$ (right)
averaged in the $Q^2$ range from 2.4 to 51.2~GeV$^2$ as functions of $x$.
The lines show the QED predictions from \cite{AZIM}.
The points are plotted at the $x$ values where the QED prediction is equal
to its mean value over the $x$ bin. }
\label{corvsx}
\end{figure}


\newpage
\begin{thebibliography}{xxx}
\bibitem{CELLO} H.J.~Behrend et al. (CELLO Collab.), Phys. Lett. {\bf B126}
(1983) 384.
\bibitem{TPC} M.P.~Cain et al. (TPC/2$\gamma$ Collab.), Phys. Lett. {\bf B147}
(1984) 232.
\bibitem{PLUTO} Ch.~Berger et al. (PLUTO Collab.), Z. Phys. {\bf C27} (1985) 249.
\bibitem{JADE} W.~Bartel et al. (JADE Collab.), Z. Phys. {\bf C30} (1986) 545.
\bibitem{MARKJ} B.~Adeva et al. (MARK J coolab.), Phys. Rev. {\bf D38} (1988)
 2665.
\bibitem{CELLO1} H.J.~Behrend et al. (CELLO Collab.), Z. Phys. {\bf C43} (1989) 1.
\bibitem{MARKII} M.~Petradza et al. (MARK II Collab.), Phys. Rev. {\bf D42} (1990)
 2171.
\bibitem{HRS} M.~Petradza et al. (HRS Collab.), Phys. Rev. {\bf D42} (1990) 2180.
\bibitem{AMY} Y.H.~Ho et al. (AMY Collab.), Phys. Lett. {\bf B244} (1990) 573.
\bibitem{OP1} R.~Akers et al. (OPAL Collab.), Z. Phys. {\bf C60} (1993) 593.
\bibitem{Poz} R.~Abreu et al. (DELPHI Collab.), Z. Phys. {\bf C69} (1996) 223.
\bibitem{OP2} K.~Ackerstaff et al. (OPAL Collab.), Z. Phys. {\bf C74} (1997) 49.
\bibitem{L3} M.~Acciarri et al. (L3 Collab.), Phys. Lett. {\bf B438} (1998) 363.
\bibitem{OP3} G.~Abbiendi et al. (OPAL Collab.), Eur. Phys. J. {\bf C11} (1999) 409.
\bibitem{Budnev} V.M.~Budnev et al., Phys. Rep. {\bf 15} (1974) 181.
\bibitem{LEP2} P.~Aurenche et al., `Physics at LEP2', eds. G.~Altarelli,
 T.~Sj\"{o}strand and F.~Zwirner, CERN 96-01 (1996) Vol.1 p.291.
\bibitem{Zerwas} C.~Peterson, P.M.~Zerwas and T.F.~Walsh, Nucl. Phys. 
 {\bf B229} (1983) 301.
\bibitem{AZIM} S.~Ong and P.~Kessler, Mod. Phys. Lett. {\bf A2} (1987) 683.\\
   Note that in the expression for the helicity terms
$I_{++,++} + I_{++,--}$ a factor $\beta^2$ is missing in front of
$(1-u^2)$ and in the expression for $I_{++,00}$ there should be the
first degree power in the denominator.
\bibitem{DELPHI} P.~Aarnio et al. (DELPHI Collab.), Nucl. Inst. Meth. {\bf A303}
(1991) 233.
\bibitem{PERF}  P.~Abreu et al. (DELPHI Collab.), Nucl. Inst. Meth. {\bf A378}
(1996) 57.
\bibitem{BDKRC} F.A.~Berends, P.H.~Daverveldt, R.~Kleiss, Comp. Phys.
 Comm. {\bf 40} (1986) 271.
\bibitem{DIAG36} F.A.~Berends, P.H.~Daverveldt, R.~Kleiss, Comp. Phys.
 Comm. {\bf 40} (1986) 285.
\bibitem{TWOGAM} S.~Nova, A.~Olshevski and T.~Todorov, 
`Physics at LEP2', CERN 96-01 (1996) Vol.2 p.224; updated version described in
`Reports of the Working Groups on Precision Calculations for LEP2 Physics', 
CERN 2000-009 (2000) 243.
\bibitem{DYMU3} J.E.~Campagne and R.~Zitoun, Z. Phys. {\bf C43} (1989) 469.
\bibitem{KORALZ} S.~Jadach et al., Comp. Phys. Comm. {\bf 79} (1994) 503.
\bibitem{pipi} Ch.~Berger et al. (PLUTO Collab.), Phys. Lett. {\bf B137} (1984)
 267.
\bibitem{offmom} P.~Abreu et al. (DELPHI Collab.), Phys. Lett. {\bf B342} (1995)
 402.
\bibitem{Berger} C.~Berger and W.~Wagner, Phys. Rep. {\bf 146} (1987) 1.
\bibitem{Agostini} G.~D'Agostini, Nucl. Inst. Meth. {\bf A362} (1995) 487.
\end{thebibliography}
\end{document}